\documentclass[fleqn,10pt]{wlscirep}
\usepackage[utf8]{inputenc}
\usepackage[T1]{fontenc}
\usepackage{bm}
\usepackage{braket}
\usepackage{booktabs, longtable, lscape, xcolor}
\usepackage{multirow, siunitx}
\usepackage[version=4]{mhchem}

\title{Resource Estimation for VQE on Small Molecules: Impact of Fermion Mappings and Hamiltonian Reductions}

\author[1]{\small Anurag K. S. V.}
\author[1]{\small Ashish Kumar Patra}
\author[1, 2]{\small Vikas Dattatraya Ghevade}
\author[1]{\small Sai Shankar P.}
\author[1, 3]{\small Ruchika Bhat}
\author[4]{\small Raghavendra V.}
\author[2]{\small Rahul Maitra}
\author[, 1]{\small Jaiganesh G.\thanks{email: jaiganesh@qclairvoyance.in, drjaiganesh15@gmail.com (Corresponding Author)}}
\affil[1]{\small Qclairvoyance Quantum Labs, Secunderabad, TG 500094, India.}
\affil[2]{\small Indian Institute of Technology Bombay, Mumbai, MH 400076, India.}
\affil[3]{\small The University of Arizona, Tucson, AZ 85721, USA.}
\affil[4]{\small SRM Institute of Science and Technology, Chennai, TN 603203, India.}

\begin{abstract}
Accurate determination of ground-state energies for molecules remains a challenge in quantum chemistry and a cornerstone for progress in fields such as drug discovery and materials design. The Variational Quantum Eigensolver (VQE) represents a leading hybrid quantum-classical paradigm for addressing this challenge; however, its widespread realization is limited by noise and the restricted scalability of current quantum hardware. Achieving efficient simulations on Noisy Intermediate-Scale Quantum (NISQ) devices and forthcoming Fault-Tolerant Application-Scalable Quantum (FASQ) systems demands a detailed understanding of how computational resources scale with molecular complexity and fermion-to-qubit encoding schemes. In this study, resource requirements for VQE implementations employing the Unitary Coupled Cluster Singles and Doubles (UCCSD) ansatz are systematically analyzed. The molecular Hamiltonian is formulated in second quantization and mapped to qubit operators through the Jordan-Wigner (JW), Bravyi-Kitaev (BK), and Parity (Pa) transformations. Hamiltonian reduction strategies, including $\mathbb{Z}_2$ tapering and frozen-core approximations, are examined to assess their effect on quantum resource scaling. The analysis reveals that appropriate transformations, when combined with symmetry-based reductions, can substantially reduce qubit counts by up to $\approx 50\%$ and quantum gate counts by up to $\approx 27.5\times$ and Hamiltonian Pauli string counts by up to $\approx 2.75\times$, relative to the corresponding unreduced Hamiltonian representations within the same active-space configuration for the representative set of molecular systems under study. These findings provide practical circuit-level insights for executing chemically relevant simulations on NISQ hardware, while establishing physical-resource baselines that may inform future logical-level analyses targeting FASQ systems.
\end{abstract}
\begin{document}

\flushbottom
\maketitle

\thispagestyle{empty}


\newpage

\section{Introduction}
Quantum computing (QC) is poised to transform computational paradigms by exploiting quantum-mechanical phenomena such as quantum superposition~\cite{Dirac1958, Feynman1982}, quantum interference~\cite{Tonomura1989, Zeilinger1999}, and quantum entanglement~\cite{Einstein1935, Horodecki2009} to perform computation using qubits rather than classical bits~\cite{Feynman1982, Shor1994, Grover1996, Shor1997, Nielsen2000}. Several classes of quantum algorithms have demonstrated theoretical promise in addressing problems in simulation~\cite{Feynman1982, Lloyd1996, AspuruGuzik2005}, optimization~\cite{Harrow2009, Farhi2014}, and search~\cite{Grover1996, Boyer1998}. Among the most compelling early applications is quantum-assisted molecular modelling, where quantum computers are employed to solve the electronic structure problem~\cite{Cao2019, McArdle2020, Patra2025} by evaluating the expectation value of the molecular Hamiltonian~\cite{McArdle2020, Bauer2020} acting on the molecular wavefunction.

The Quantum Phase Estimation (QPE) algorithm provides an asymptotically optimal, fully quantum method for extracting the eigenvalues of the Hamiltonian by coherently simulating its time evolution and performing an inverse quantum Fourier transform-based phase readout~\cite{AspuruGuzik2005, OMalley2016}. Although QPE enables chemically accurate energy estimation with polynomial scaling, its implementation requires deep circuits, long coherence times, and fault-tolerant error correction. These requirements limit its near-term applicability on NISQ~\cite{Preskill2018} hardware and underscore the long-term need for FASQ~\cite{Eisert2025} devices. The VQE algorithm, on the other hand, approximates the ground-state energy of the given Hamiltonian by preparing parameterized quantum circuits (ansätze) and iteratively optimizing their parameters using classical optimizer feedback loops~\cite{Peruzzo2014, McClean2016}. A widely adopted chemistry-inspired ansatz is the UCCSD ansatz~\cite{Barkoutsos2018}, which provides a physically motivated way to represent correlated electronic wavefunctions on QC hardware. The practical realization of VQE with UCCSD remains strongly constrained by the current state of QC hardware, which is dominated by noise, limited connectivity, and shallow circuit depths characteristic of the NISQ era~\cite{Tilly2022}.

Given these engineering challenges and the limited hardware availability of NISQ and the long-term development of FASQ devices, a priori resource estimation is essential for assessing the scale of quantum circuits that can be realistically executed. Recent research has examined the algorithmic components that fundamentally determine the resource demands of quantum chemistry simulations, particularly for small‐molecule VQE workflows. Foundational comparisons of fermion-to-qubit mappings (FTQMs) across 86 molecular systems have shown that the BK transformation often reduces Pauli string lengths and gate counts relative to the JW mapping, offering tangible savings for NISQ hardware implementations~\cite{Tranter2018}. Complementary investigations of UCCSD-VQE accuracy on small molecules demonstrated that the method can capture chemically meaningful energies for both open- and closed-shell species, while also highlighting that practical usefulness depends on the attainable qubit counts, circuit depths, and noise levels of NISQ devices~\cite{Michael2019}. Beyond chemistry problems, broader VQE resource studies, such as those for the Hubbard model, have underscored the scaling bottlenecks associated with two-qubit operations and measurement shot requirements, showing that meaningful simulations of 50-qubits already require tens of thousands of entangling gates and low two-qubit-gate-error rates ($\sim10^{-4}$)~\cite{Cai2020}. These insights collectively emphasize that, even before algorithmic optimization, Hamiltonian structure, mapping choice, and ansatz construction are major determinants of VQE feasibility.

Further work has focused on quantifying and reducing the dominant runtime overheads associated with measurement, optimization, and Hamiltonian evaluation. Robust amplitude-estimation-based measurement protocols have been shown to reduce the shot complexity of VQE energy estimation by one to two orders of magnitude compared with standard sampling strategies, though they remain impractically demanding for many chemical systems~\cite{Johnson2022}. Parallel efforts have produced resource-estimation tools, such as QREChem, which analyze logical gate counts for Trotterized QPE and provide baselines for molecules ranging from $10^{7}$ to $10^{15}$ $T$ gates, illustrating the enormous gap between NISQ and FASQ requirements~\cite{Otten2023}. The measurement overhead associated with evaluating the expectation value of the Hamiltonian in VQE represents a well-recognised practical bottleneck~\cite{Wecker2015, Gonthier2022}. The number of required measurement circuits scales with the number of unique Pauli strings in the Hamiltonian, and achieving chemically meaningful precision requires large numbers of circuit repetitions, a cost that grows with system size. Reviews on quantum-chemistry measurement strategies for VQE and QPE~\cite{Patel2025}, as well as practical hardware studies evaluating UCCSD versus hardware-efficient ansätze on IBM Quantum hardware~\cite{Belaloui2025}, further highlight that measurement grouping, symmetry exploitation, and noise-robust ansatz choices are essential for molecule simulations on QC hardware. More broadly, recent assessments of quantum advantage in computational chemistry~\cite{Hans2025} and perspectives on 25 to 100 logical-qubit early FASQ devices~\cite{Alexeev2025} suggest that quantum chemistry will remain a leading candidate for early utility, but only with careful co-design of resource-aware algorithms and chemically relevant problem instances. Together, these works establish resource estimation not only as a forecasting tool but also as a design principle, central to evaluating the realistic prospects of QC simulations for small-molecule electronic structure calculations.

In this work, we perform a systematic, hardware-agnostic quantum-resource analysis of VQE with the UCCSD ansatz for a representative set of small molecules: $H_{2}$, $LiH$, $HF$, $BeH_2$, $H_2O$, $N_{2}$, $O_{2}$, $CO$, $NH_{3}$, $CH_{4}$, $C_{2}H_{2}$, $H_{2}O_{2}$, and $C_{2}H_{4}$, on QC simulators, thereby isolating mapping-dependent effects from hardware-specific compilation constraints. To enable a balanced and consistent comparison across fermion-to-qubit mappings, we adopt the UCCSD ansatz as a static, chemically motivated, and systematically improvable reference framework that remains independent of hardware topology. Hardware-efficient ansatz (HEA)~\cite{Kandala2017, Leone2024} and adaptive ansatz (ADAPT)~\cite{Grimsley2019, Nonia2025} constructions can significantly reduce circuit depth relative to UCCSD. However, these adaptive schemes require substantial additional measurement overhead during ansatz construction. Recent developments, including measurement-efficient ansatz strategies such as COMPASS~\cite{Mondal2023} and COMPASS-PRO~\cite{Mondal2026}, SURGE-VQE~\cite{Dipanjali2025}, as well as approaches such as COMPACT~\cite{Dipanjali2024} and RBM-VQE~\cite{Halder2024, Halder2025} have partially or fully alleviated these costs. While UCCSD is known to generate deeper circuits than hardware-efficient, adaptive, and measurement-efficient alternatives, the resulting circuit metrics may be interpreted as representative upper bounds on resource requirements within the single- and double-excitation manifolds of a chemically motivated ansatz framework. A systematic evaluation of hardware-efficient, adaptive, and measurement-efficient ansatz families is therefore outside the scope of the present study.

The resource metrics reported here are physical circuit-level quantities after transpilation, directly relevant to NISQ and near-term quantum hardware. Logical-level resource estimates, including logical qubit counts, $T$-gate counts, and error-correction overheads, as well as full measurement-cost analysis, including shot-count and Pauli grouping overheads, are outside the scope of this work but represent a natural extension of the baselines established here. Concretely, we (1) quantify qubit counts, Pauli string counts, and quantum gate counts, across common FTQMs~\cite{McArdle2020}, and (2) evaluate the impact of the frozen-core approximation~\cite{Romero2019} and $\mathbb{Z}_2$-symmetry-based qubit tapering~\cite{Bravyi2017} on reducing resource requirements. The theoretical background of the work is outlined in Section~\ref{sec2:theory}, followed by the methodology employed in Section~\ref{sec3:methodology}. The results and their discussion are presented in Section~\ref{sec4:results}, and the concluding remarks are provided in Section~\ref{sec5:conclusion}.

\section{Theoretical Background} \label{sec2:theory}
Accurate quantum simulation of molecular systems requires a rigorous formulation of the molecular Hamiltonian and its associated wavefunction, followed by a sequence of transformations that render the problem suitable for execution on quantum hardware. Starting from the non-relativistic molecular Hamiltonian $\hat{H}_{mol}$ and the molecular wavefunction $\psi(\vec{r}, \vec{R})$ (Section~\ref{sec2.1:molecular_hamiltonian}), the problem is systematically reduced to an effective electronic Hamiltonian $\hat{H}_{elec}$ and electronic wavefunction $\psi_{e}(\vec{r}; \vec{R})$ (Section~\ref{sec2.2:electronic_hamiltonian}). The Hamiltonian $\hat{H}_{elec}$ is then expressed in its second-quantized fermionic form $\hat{H}_{sq}$ with the corresponding state $\ket{\psi_{sq}}$ (Section~\ref{sec2.3:secondquantized_hamiltonian}). Subsequently, $\hat{H}_{sq}$ is mapped to a qubit Hamiltonian $\hat{H}_{qubit}$, and the state $\ket{\psi_{sq}}$ is transformed to the qubit basis to obtain $\ket{\Psi_{qubit}}$ (Section~\ref{sec2.4:qubit_hamiltonian}) using FTQMs such as JW~\cite{Jordan1928, Nielsen2005}, BK~\cite{Bravyi2002, Seeley2012, Tranter2015}, and Pa~\cite{Seeley2012, McArdle2020} (Section~\ref{sec2.5:ftqm}). This enables quantum simulation of the molecular system on universal gate-based quantum computers. To further enhance simulation efficiency, Hamiltonian reduction techniques, such as freezing core orbitals to obtain $\hat{H}_{fc}$ and tapering symmetries to derive the symmetry-reduced Hamiltonian $\hat{H}_{tapered}$, may be applied (Section~\ref{sec2.6:hamiltonian_reduction}).

\subsection{Non-relativistic Molecular Hamiltonian and Wavefunction}
\label{sec2.1:molecular_hamiltonian}
Within the framework of non-relativistic quantum mechanics~\cite{szabo1996ch2}, the total energy of a molecular system is represented by the Hamiltonian operator, which can be expressed as the sum of its kinetic energy and potential energy:
\begin{equation}\label{eq:H_total}
\hat{H} = \hat{T} + \hat{V},
\end{equation}
where the total kinetic energy operator $\hat{T}$ in Eq.~(\ref{eq:H_total}) consists of the nuclear $\hat{T}_{N}$ and the electronic $\hat{T}_{e}$ contributions:
\begin{equation}\label{eq:T_components}
\hat{T} = \hat{T}_N + \hat{T}_e;
\quad
\hat{T}_{N} = - \sum_{A=1}^{N_n} \frac{\hslash^{2}}{2 M_A} \hat{\nabla}_A^2,
\quad
\hat{T}_{e} = - \sum_{i=1}^{N_e} \frac{\hslash^{2}}{2 m_{e}} \hat{\nabla}_i^2, 
\end{equation}
and the potential energy operator $\hat{V}$ in Eq.~(\ref{eq:H_total}) consists of all the Coulomb interactions present in the molecular system. These consist of the electron–nuclear ($\hat{V}_{eN}$), electron–electron ($\hat{V}_{ee}$), and nuclear–nuclear ($\hat{V}_{NN}$) interactions:
\begin{equation}\label{eq:V_components}
\hat{V} = \hat{V}_{eN} + \hat{V}_{ee} + \hat{V}_{NN};
\quad
\hat{V}_{eN} = - \sum_{i=1}^{N_e} \sum_{A=1}^{N_n} \frac{Z_A \cdot e^2}{|\vec{r}_{iA}|},
\quad
\hat{V}_{ee} = \frac{1}{2}\sum^{N_e}_{\substack{i,j\\ i\neq j}} \frac{e^2}{|\vec{r}_{ij}|},
\quad
\hat{V}_{NN} = \frac{1}{2}\sum_{\substack{A, B \\ A \neq B}}^{N_n} \frac{Z_A \cdot Z_B \cdot e^2}{|\vec{R}_{AB}|},
\end{equation} 
combining these components from Eqs.~(\ref{eq:T_components}) and (\ref{eq:V_components}) yields the standard non-relativistic molecular Hamiltonian $\hat{H}_{mol}$ in atomic units~\cite{szabo1996ch2}:
\begin{equation}\label{eq:molecular_hamiltonian}
\hat{H}_{mol} = 
- \sum_{A=1}^{N_n} \frac{1}{2 M_A} \hat{\nabla}_A^2
- \sum_{i=1}^{N_e} \frac{1}{2} \hat{\nabla}_i^2
- \sum_{i=1}^{N_e} \sum_{A=1}^{N_n} \frac{Z_A}{|\vec{r}_{iA}|}
+ \frac{1}{2}\sum^{N_e}_{\substack{i,j\\ i\neq j}} \frac{1}{|\vec{r}_{ij}|}
+ \frac{1}{2}\sum_{\substack{A, B \\ A \neq B}}^{N_n} \frac{Z_A \cdot Z_B}{|\vec{R}_{AB}|},
\end{equation}

\begin{figure}[ht]
\centering
    \includegraphics[width=0.6\linewidth]{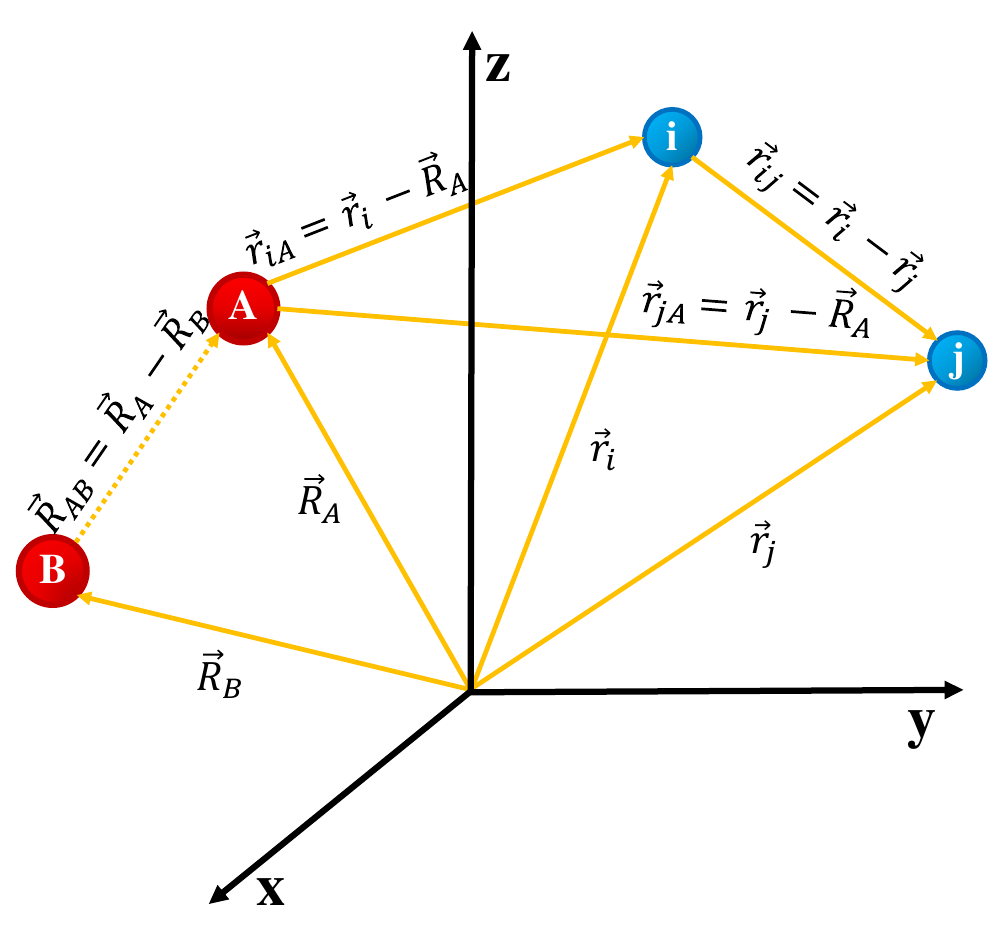}
    \caption{Schematic representation of a diatomic molecular system described by the non-relativistic molecular Hamiltonian in Eq.~(\ref{eq:molecular_hamiltonian}). The nuclei \( A \) and \( B \) (shown in red) and electrons \( i \) and \( j \) (shown in blue) are located by their respective position vectors \( \vec{R}_A, \vec{R}_B, \vec{r}_i, \) and \( \vec{r}_j \) with respect to the origin. The inter-particle vectors \( \vec{r}_{iA}, \vec{r}_{jA}, \) and \( \vec{r}_{ij} \) are represented as solid lines, while the internuclear vector \( \vec{R}_{AB} = \vec{R}_A - \vec{R}_B \) is shown as a dotted line to indicate that, under the Born--Oppenheimer approximation, the internuclear separation is treated as a fixed parameter.}
    \label{fig:molecular-geometry}
\end{figure}

In Eqs.~(\ref{eq:T_components}), (\ref{eq:V_components}), and (\ref{eq:molecular_hamiltonian}), $\hslash = {h}/{2\pi}$, where $h$ is Planck’s constant.  $\hat{\nabla}_A^2$ and $\hat{\nabla}_i^2$ denote the Laplacian operators corresponding to nucleus $A$ and electron $i$, respectively. $N_e$ and $N_n$ represent the numbers of electrons and nuclei; $Z_A$($Z_B$) and $M_A$ correspond to the atomic number and mass of nucleus $A$($B$); $m_e$ is the mass of an electron; and $e$ is the elementary charge. $|\vec{r}_{iA}|$, $|\vec{r}_{ij}|$, and $|\vec{R}_{AB}|$ denote the electron-nucleus, electron-electron, and nucleus-nucleus separations, respectively, as shown in Figure~\ref{fig:molecular-geometry}. In Eq.~(\ref{eq:molecular_hamiltonian}), the first two terms describe the nuclear and electronic kinetic energy operators ($\hat{T}_N$ and $\hat{T}_e$ ), the third term represents the electron–nucleus attraction ($\hat{V}_{eN}$), the fourth term corresponds to the electron–electron repulsion ($\hat{V}_{ee}$), and the final term accounts for the nucleus–nucleus repulsion ($\hat{V}_{NN}$).

The total molecular wavefunction $\psi(\vec{r}, \vec{R})$ is defined over both the electronic coordinates $\vec{r} = (\vec{r}_1, \vec{r}_2, \ldots, \vec{r}_{N_e})$ and the nuclear coordinates $\vec{R} = (\vec{R}_1, \vec{R}_2, \ldots, \vec{R}_{N_n})$~\cite{szabo1996ch2, Helgaker2000}:
\begin{equation}
    \psi(\vec{r}, \vec{R}) = \psi(\vec{r}_1, \vec{r}_2, \ldots, \vec{r}_{N_e}, \vec{R}_1, \vec{R}_2, \ldots, \vec{R}_{N_n}).
\end{equation}

This wavefunction encompasses all dynamical variables of the system, including the correlated motion of electrons and nuclei, within the Born-Huang representation~\cite{Born1954}. 

Accordingly, the molecular energy $E_{mol}$ is obtained by solving the time-independent molecular Schrödinger equation:
\begin{equation}\label{eq:mol_schrodinger}
\hat{H}_{mol} \, \psi(\vec{r}, \vec{R}) = E_{mol} \, \psi(\vec{r}, \vec{R}),
\end{equation}
where $\hat{H}_{mol}$ is the non-relativistic molecular Hamiltonian defined in Eq.~(\ref{eq:molecular_hamiltonian}). The molecular wavefunction $\psi(\vec{r}, \vec{R})$ is antisymmetric under exchange of any two electrons, consistent with the Pauli exclusion principle~\cite{szabo1996ch2, Helgaker2000, Fetter2003ch1, Shavitt2009ch2}. 

This formulation, referred to as the first-quantized molecular Hamiltonian, provides the foundation for further approximations such as the Born-Oppenheimer (B.O.) separation of electronic and nuclear motion, leading to the first-quantized electronic Hamiltonian $\hat{H}_{elec}$~\cite{Born1927}. While the B.O. separation provides a practical and widely used route to electronic-structure calculations, several non-B.O. frameworks, such as the early protonic-structure~\cite{Thomas1969pr, Thomas1969cpl}, nuclear-orbital plus molecular-orbital (NOMO)~\cite{Tachikawa1998, Hiromi2001, Hiromi2002, Hiromi2003, Hiromi2007}, multicomponent molecular-orbital (MCMO)~\cite{Tachikawa2001, Takayoshi2009}, nuclear-electronic orbital (NEO)~\cite{Webb2002, Pavosevic2020}, and more recent constrained NEO (CNEO)~\cite{Xu2020, Xi2020, Culpitt2025}, treat electrons and selected nuclei quantum mechanically. A detailed treatment of these non-B.O. formulations, although essential for fully correlated nuclear-electronic dynamics, lies beyond the scope of the present work. Here, we focus on $\hat{H}_{mol}$ derived under the B.O. approximation, yielding $\hat{H}_{elec}$ as discussed in the next section.

\subsection{Electronic Hamiltonian and Wavefunction}
\label{sec2.2:electronic_hamiltonian}

To simplify the many-body system in Eq.~(\ref{eq:molecular_hamiltonian}), the B.O. approximation assumes that nuclei remain effectively stationary due to their much larger masses compared to electrons. Consequently, the nuclear kinetic energy term $\hat{T}_N$ can be neglected, and the nuclei are treated as fixed classical point charges~\cite{Born1927, Born1954, szabo1996ch2, Helgaker2000}. 

The resulting first-quantized electronic Hamiltonian $\hat{H}_{elec}$, parametrized by the nuclear coordinates $\vec{R}$, is expressed as:
\begin{equation}\label{eq:electronic_hamiltonian}
\hat{H}_{elec} =
- \frac{1}{2} \sum_{i=1}^{N_e} \hat{\nabla}_i^2
- \sum_{i=1}^{N_e} \sum_{A=1}^{N_n} \frac{Z_A}{|\vec{r}_{iA}|}
+ \frac{1}{2}\sum_{\substack{i,j \\ i \neq j}}^{N_e} \frac{1}{|\vec{r}_{ij}|}
+ \hat{V}_{NN},
\end{equation}
where $\hat{V}_{NN} = \frac{1}{2}\sum_{\substack{A,B \\ A \neq B}}^{N_n} \frac{Z_A Z_B}{|\vec{R}_{AB}|}$ represents the constant nuclear–nuclear repulsion energy. 

The corresponding electronic wavefunction depends explicitly on the electronic coordinates $\vec{r} = (\vec{r}_1, \vec{r}_2, \ldots, \vec{r}_{N_e})$ and parametrically on the nuclear coordinates $\vec{R}$. Incorporating spin, each electron is described by combined space–spin coordinates $\chi_i = (\vec{r}_i, \sigma_i)$, where $\sigma_i \in \{\uparrow, \downarrow\}$. The complete electronic wavefunction $\psi_e(\chi;\vec{R})$ is therefore given by:
\begin{equation}
    \psi_e(\chi;\vec{R}) = \psi_e(\chi_1, \chi_2, \ldots, \chi_{N_e}; \vec{R}).
\end{equation}

The electronic  energy at a fixed nuclear configuration $E_{elec}(\vec{R})$ is obtained by solving the time-independent electronic Schrödinger equation:
\begin{equation}\label{eq:electronic_schrodinger}
\hat{H}_{elec}(\vec{R}) \, \psi_e(\chi; \vec{R}) = E_{elec}(\vec{R}) \, \psi_e(\chi; \vec{R}),
\end{equation}
where $\hat{H}_{elec}(\vec{R})$ is the electronic Hamiltonian for a given nuclear geometry $\vec{R}$.

However, operating directly in the continuous first-quantized coordinate representation of $\hat{H}_{elec}$ and $\psi_e(\chi; \vec{R})$ is computationally inefficient for many-body systems. To enable efficient numerical treatment and facilitate quantum simulation, the Hamiltonian is reformulated in second quantization as $\hat{H}_{sq}$, where the wavefunction of the system is expressed in the discrete occupation-number basis as $\ket{\psi_{sq}}$. In this representation, the Hamiltonian is written in terms of fermionic creation and annihilation operators ($\hat{a}^\dagger, \hat{a}$), as discussed in detail in the following section.

\subsection{Second Quantization: Fermionic Representation}
\label{sec2.3:secondquantized_hamiltonian}

In the second-quantized framework, the first-quantized electronic Hamiltonian $\hat{H}_{elec}$ is expressed as:
\begin{equation}\label{eq:second_quantized_hamiltonian}
\hat{H}_{sq} = \sum_{pq} h_{pq}\,\hat{a}^\dagger_p \hat{a}_q
+ \frac{1}{2}\sum_{pqrs} h_{pqrs}\,\hat{a}^\dagger_p \hat{a}^\dagger_q \hat{a}_r \hat{a}_s,
\end{equation}
where $\hat{a}^\dagger_p$ and $\hat{a}_q$ are the $(\hat{a}^{\dagger}, \hat{a})$ acting on spin orbitals labeled by indices $p$ and $q$. These operators obey the canonical anti-commutation relations~\cite{szabo1996ch2, Helgaker2000}:
\begin{equation}
     \{\hat{a}^\dagger_p, \hat{a}_q\} = \delta_{pq},
     \quad
     \{\hat{a}_p, \hat{a}_q\} = 0,
     \quad
     \{\hat{a}^\dagger_p, \hat{a}^\dagger_q\} = 0.
\end{equation}
The coefficients $h_{pq}$ and $h_{pqrs}$ represent the one- and two-electron integrals evaluated over the selected spin orbital basis. The one-electron integrals are given by:
\begin{equation}\label{eq:one_electron_integral} h_{pq} = \int d\vec{r}\, \xi_p^*(\vec{r}) \Biggl[-\frac{1}{2}\hat{\nabla}^2 - \sum_{A=1}^{N_n} \frac{Z_A}{|\vec{r} - \vec{R}_A|}\Biggr] \chi_q(\vec{r}),
\end{equation}
incorporate the electronic kinetic energy and electron–nucleus attraction, where $\xi_p(\vec{r})$ is the spin orbital for index $p$, $Z_A$ represents the atomic number associated with $A$, $\vec{R}_A$ is its position. The two-electron integrals:
\begin{equation}\label{eq:two_electron_integral}
h_{pqrs} = \int \int d\vec{r}_1 d\vec{r}_2 \frac{\xi_p^*(\vec{r}_1) \xi_q^*(\vec{r}_2) \xi_r(\vec{r}_1) \xi_s(\vec{r}_2)}{|\vec{r}_1 - \vec{r}_2|},
\end{equation}
account for electron-electron repulsion. The prefactor $1/2$ in Eq.~\eqref{eq:second_quantized_hamiltonian} avoids double counting over electron pairs.

The indices $p, q, r, s$ span the complete set of spin orbitals in the chosen basis (e.g., STO-3G~\cite{Hehre1969}, cc-pVDZ~\cite{Dunning1989}). The resulting operator $\hat{H}_{sq}$ acts on Fock space~\cite{Fock1932}, a discrete Hilbert space spanned by the occupation-number basis:
\begin{equation} \label{eq:occ-no_basis}
    \ket{n_{m-1}} \otimes \ket{n_{m-2}} \otimes \ldots \otimes \ket{n_{1}} \otimes \ket{n_{0}},
    \qquad
    n_j \in \{0,1\},
\end{equation} 
where $m$ denotes the total number of spin orbitals, and $\ket{n_j}$ denotes whether the $j^{th}$ spin orbital is occupied. The many-electron wavefunction in its second quantized form $\ket{\psi_{sq}}$ is then expressed as a superposition of these occupation-number basis states:
\begin{equation}
    \ket{\psi_{sq}} = \sum_j c_j \bigotimes_{k=0}^{m-1} \ket{n_{k}} = \sum_{j} c_j \ket{\phi_j},\quad n_k \in \{0, 1\},
\end{equation}
where $c_j$ are the probability amplitudes of the basis states $\ket{\phi_j}$ expressed in occupation number basis (Eq.~(\ref{eq:occ-no_basis})). The time-independent electronic Schrödinger equation in occupation-number basis is given as:
\begin{equation}
    \hat{H}_{sq}(\vec{R}) \ket{\psi_{sq}} = E_{elec}(\vec{R}) \ket{\psi_{sq}}
\end{equation}

where, $H_{sq}(\vec{R})$ is second quantized electronic Hamiltonian for the given nuclear geometry $\vec{R}$.

This fermionic representation $\hat{H}_{sq}$ forms the bridge between $\hat{H}_{elec}$ and its qubit-based encoding via mappings that transform fermionic operators into qubit operators suitable for implementation on QC hardware. The resulting qubit Hamiltonian $\hat{H}_{qubit}$ serves as the input to quantum algorithms, discussed in the next section.

\subsection{Qubit Hamiltonian and Wavefunction}
\label{sec2.4:qubit_hamiltonian}

The second-quantized Hamiltonian, $\hat{H}_{sq}$, can subsequently be transformed into the $\hat{H}_{qubit}$, represented as a linear combination of tensor products of Pauli operators ($\hat{\sigma}_{x}$, $\hat{\sigma}_{y}$, $\hat{\sigma}_{z}$) and the identity operator $\hat{\mathbb{I}}_{2}$:
\begin{equation}
\hat{\sigma}_{x} =
\begin{bmatrix}
0 & 1 \\
1 & 0
\end{bmatrix}, 
\quad
\hat{\sigma}_{y} =
\begin{bmatrix}
0 & -i \\
i & 0
\end{bmatrix}, 
\quad
\hat{\sigma}_{z} =
\begin{bmatrix}
1 & 0 \\
0 & -1
\end{bmatrix},
\quad
\hat{\mathbb{I}}_{2} =
\begin{bmatrix}
1 & 0 \\
0 & 1
\end{bmatrix}.
\end{equation}
This transformation is performed through FTQMs such as JW, BK, and Pa. These mappings yield the $\hat{H}_{qubit}$:
\begin{equation}
    \label{eq:qubit_hamiltonian}
    \hat{H}_{qubit} = \sum_{j} {\tau}_j \bigotimes_{k=0}^{m-1}\hat{\sigma}^k_p,\quad \hat{\sigma}_p \in \{\hat{\sigma}_x, \hat{\sigma}_y, \hat{\sigma}_z, \hat{\mathbb{I}}_2\},
\end{equation}
where $\bigotimes_{k=0}^{m-1} \hat{\sigma}_p^k$ are the Pauli strings of $\hat{H}_{qubit}$ and $\tau_j$ are the co-efficients of the Pauli strings. The corresponding statevector representation of the wavefunction $\ket{\Psi_{qubit}}$ takes the form:
\begin{equation}
    \ket{\Psi_{qubit}} = \sum_{j} c_j \bigotimes_{k=0}^{m-1} \ket{q_k} = \sum_j c_j \ket{\Phi_j},\quad q_k \in \{0,1\},
\end{equation}
where, $c_j$ are the probability amplitudes corresponding to the computational basis states $\ket{\Phi_j}$. $\ket{\Phi_j}$ is given as $\bigotimes_{k=0}^{m-1} \ket{q_k}$, and $\ket{q_k}$ denotes the state of the $k^{th}$ qubit. Therefore, the general time-independent electronic Schrödinger equation given $\hat{H}_{qubit}$ and $\ket{\Psi_{qubit}}$ is expressed as:
\begin{equation}
    \hat{H}_{qubit}(\vec{R}) \ket{\Psi_{qubit}} = E_{elec}(\vec{R}) \ket{\Psi_{qubit}},
\end{equation}
where, $\hat{H}_{qubit}(\vec{R})$ is the qubit Hamiltonian for a given nuclear geometry $\vec{R}$. Given an FTQM, the representations of $\hat{H}_{qubit}$  and $\ket{\Psi_{qubit}}$ are detailed in the next section, which serve as the input to quantum algorithms such as VQE and QPE, applicable to both near-term and future fault-tolerant architectures.

\subsection{Fermion-to-Qubit Mappings (FTQMs)} \label{sec2.5:ftqm}
Several mapping schemes have been proposed to achieve FTQM efficiently, each balancing trade-offs between operator locality, qubit overhead, and circuit depth~\cite{Jordan1928, Bravyi2002,  Seeley2012, Steudtner2018, Jiang2020, Derby2021, Miller2023, OBrien2024, Parella2024, Yu2025, Liu2025}. In this section, we outline the theoretical background of the three most widely used mappings: JW~\cite{Jordan1928}, BK~\cite{Bravyi2002}, and Pa~\cite{Seeley2012}. For each mapping, we present the corresponding basis construction and show how $(\hat{a}^{\dagger}, \hat{a})$ are represented in terms of Pauli operators.

\subsubsection{Jordan–Wigner (JW)}
In the JW mapping, each qubit represents the occupancy of an individual spin orbital, with the computational states $\ket{0}$ and $\ket{1}$ denoting unoccupied and occupied orbitals, respectively~\cite{McArdle2020}. This formulation aligns with the occupation-number basis and is often referred to as the JW basis:
\begin{equation} \ket{n_{m-1}} \otimes \ket{n_{m-2}} \otimes \ldots \otimes \ket{n_{1}} \otimes \ket{n_{0}} \rightarrow
\ket{q_{m-1}} \otimes \ket{q_{m-2}} \otimes \cdots \otimes \ket{q_1}\otimes \ket{q_0}, \qquad\ q_j = f_j \in \{0,1\}.
\end{equation}
In this representation, the state of each qubit $\ket{q_j}$ encodes the occupation number $f_j$ of the $j^{\text{th}}$ spin orbital. The corresponding $(\hat{a}^{\dagger}, \hat{a})$ in the JW mapping are given by:
\begin{equation} 
    \hat{a}_j^\dagger \equiv
    \frac{1}{2} (\hat{\sigma}_{x}^{j} - i \hat{\sigma}_{y}^{j}) \otimes \hat{\sigma}_{z,\rightarrow}^{j-1}, \qquad 
    \hat{a}_j \equiv \frac{1}{2} (\hat{\sigma}_{x}^{j} + i \hat{\sigma}_{y}^{j}) \otimes \hat{\sigma}_{z,\rightarrow}^{j-1}, 
\end{equation}
where $ \hat{\sigma}_{z,\rightarrow}^{j} \equiv \hat{\sigma}_{z}^{j} \otimes \hat{\sigma}_{z}^{j-1} \otimes \cdots \otimes \hat{\sigma}_{z}^{1} \otimes \hat{\sigma}_{z}^{0},$ and any qubit not explicitly acted upon is implicitly operated on by $\hat{\mathbb{I}}_{2}$. The operator $\hat{\sigma}_{z,\rightarrow}^{j-1}$ acts as a parity string, with eigenvalues $\pm 1$ corresponding to even or odd parity states, respectively~\cite{Seeley2012}. 

While conceptually straightforward, the JW transformation introduces long Pauli strings, leading to increased circuit depth for non-local operators. The BK transformation addresses this limitation by balancing parity and occupation information across qubits.

\subsubsection{Bravyi–Kitaev (BK)}
In the BK transformation, each qubit encodes a partial parity of the fermionic occupation numbers, rather than a single orbital occupation~\cite{McArdle2020}. This results in a more balanced distribution of parity and occupation information across qubits, leading to logarithmic rather than linear operator locality~\cite{Bravyi2002}. The occupation numbers contributing to each partial sum are determined by the BK transformation matrix $\beta_{m}$:
\begin{equation} 
\ket{n_{m-1}} \otimes \ket{n_{m-2}} \otimes \ldots \otimes \ket{n_{1}} \otimes \ket{n_{0}} 
\rightarrow
\ket{q_{m-1}} \otimes \ket{q_{m-2}} \otimes \cdots \otimes \ket{q_1}\otimes \ket{q_0}, 
\qquad
q_j = \left[ \sum_{k=0}^{j} [\beta_{m}]_{jk} f_{k} \right] \bmod 2.
\end{equation}
Here, $\beta_{m}$ denotes the transformation matrix that maps occupation-number basis vectors of dimension $m$ to their corresponding BK basis representation. The matrix is recursively defined as:
\begin{equation}
    \beta_{1} = \left[ 1 \right],
    \quad
    \beta_{2^{l+1}} = 
    \begin{bmatrix}
        \beta_{2^{l}} & \alpha_{0} \\
        \alpha_{1} & \beta_{2^{l}}
    \end{bmatrix},
\end{equation}
In this formulation, $\alpha_{0}$ represents a $[2^{l} \times 2^{l}]$ zero matrix, while $\alpha_{1}$ is also a $[2^{l} \times 2^{l}]$ zero matrix except for its bottom row, which is filled with ones. The corresponding $(\hat{a}^{\dagger}, \hat{a})$ in this mapping are defined as:
\begin{equation}
        \hat{a}^\dagger_j \equiv \frac{1}{2}(\hat{\sigma}_{x}^{U(j)} \otimes \hat{\sigma}_{x}^{j} \otimes \hat{\sigma}_{z}^{P(j)} - i \hat{\sigma}_{x}^{U(j)} \otimes \hat{\sigma}_{y}^{j} \otimes \hat{\sigma}_{z}^{\rho(j)});
        \qquad
        \hat{a}_j \equiv \frac{1}{2}(\hat{\sigma}_{x}^{U(j)} \otimes \hat{\sigma}_{x}^{j} \otimes \hat{\sigma}_{z}^{P(j)} + i \hat{\sigma}_{x}^{U(j)} \otimes \hat{\sigma}_{y}^{j} \otimes \hat{\sigma}_{z}^{\rho(j)}).
\end{equation}
Where,
\begin{equation}
  \rho(j) \equiv \left\{ 
  \begin{array}{l l}
    P(j) & \quad {\rm if~}j{\rm~is~even;}\\
    R(j) & \quad {\rm if~}j{\rm~is~odd.}\\
  \end{array} \right.
\end{equation}
and $R(j) \equiv P(j) \setminus F(j)$, where $R(j)$, $P(j)$, $F(j)$, and $U(j)$ correspond to the remainder, parity, flip, and update sets, respectively~\cite{Seeley2012}.

Alternatively, the Parity mapping encodes parity information directly on each qubit, providing symmetry-related benefits and facilitating efficient qubit tapering.

\subsubsection{Parity (Pa)}
In the Parity mapping, parity information is encoded locally on each qubit, whereas occupation numbers are represented non-locally~\cite{McArdle2020}. The $j^{\text{th}}$ qubit thus corresponds to the parity of the first $j$ fermionic modes:
\begin{equation} 
\ket{n_{m-1}} \otimes \ket{n_{m-2}} \otimes \ldots \otimes \ket{n_{1}} \otimes \ket{n_{0}}
\rightarrow
\ket{q_{m-1}} \otimes \ket{q_{m-2}} \otimes \cdots \otimes \ket{q_1}\otimes \ket{q_0}, 
\quad
q_j = \left[ \sum_{k=0}^{j} [\pi_{m}]_{jk} f_{k} \right] \bmod 2 = \left[ \sum_{k=0}^{j} f_{k} \right] \bmod 2,
\end{equation}
where the associated transformation matrix $\pi_m$ is defined as:
\begin{equation}
    [\pi_m]_{jk} = \left\{ 
    \begin{array}{l l}
    1 & \quad j < k \\
    0 & \quad j \geq k \\
    \end{array} \right..
\end{equation}
The $(\hat{a}^{\dagger}, \hat{a})$ under this mapping are expressed as:
\begin{equation}
    \hat{a}_j^\dagger \equiv \frac{1}{2} (\hat{\sigma}_{x, \leftarrow}^{j+1} \otimes \hat{\sigma}_{x}^{j} \otimes \hat{\sigma}_{z}^{j-1} - i \hat{\sigma}_{x,\leftarrow}^{j+1} \otimes \hat{\sigma}_{y}^{j}); 
    \qquad
    \hat{a}_j \equiv \frac{1}{2} (\hat{\sigma}_{x,\leftarrow}^{j+1} \otimes \hat{\sigma}_{x}^{j} \otimes \hat{\sigma}_{z}^{j-1} + i \hat{\sigma}_{x,\leftarrow}^{j+1} \otimes \hat{\sigma}_{y}^j),
\end{equation}
here, $\hat{\sigma}_{x,\leftarrow}^{j} \equiv \hat{\sigma}_{x}^{m-1} \otimes \hat{\sigma}_{x}^{m-2} \otimes \cdots \otimes \hat{\sigma}_{x}^{j+1} \otimes \hat{\sigma}_{x}^{j}$ denotes the update operator, responsible for flipping all qubits associated with partial sums that contain the $(j - 1)^{\text{th}}$ orbital whenever its occupation state changes~\cite{Seeley2012}.

In Pa, when spin-orbitals are ordered in separate spin blocks, and the total particle number (and/or spin sector) is known, it permits an immediate two-qubit reduction by fixing global parity eigenvalues. This two-qubit reduction is a consequence of two independent $\mathbb{Z}_{2}$ symmetries (particle-number parity and an ordering-induced parity); each fixed symmetry allows removing one qubit by replacing its Pauli operator with its eigenvalue~\cite{Seeley2012, Bravyi2017}.

\subsection{Hamiltonian Reduction Strategies} \label{sec2.6:hamiltonian_reduction}
Following the transformation of the fermionic Hamiltonian into its qubit representation $\hat{H}_{qubit}$, additional reductions can be applied to optimize computational resources without compromising the physical fidelity of the simulation. Despite efficient FTQMs, the resulting $\hat{H}_{qubit}$ often contains redundant degrees of freedom that exceed the capabilities of current NISQ and near-term devices. To address this, reduction techniques are used to minimize the active qubit space and simplify the operator structure while preserving the eigenvalue spectrum relevant to the target state.

\subsubsection{Frozen-Core Approximation}
The frozen-core (FC) approximation leverages the observation that core orbitals that are deeply bound and chemically inert contribute negligibly to correlation effects among valence electrons. These orbitals are treated as doubly occupied and excluded from the active orbital space used in electronic correlation and qubit mapping~\cite{Roos1980, Battaglia2024}, yielding an FC Hamiltonian $\hat{H}_{fc}$:
\begin{equation}
    \hat{H}_{fc} = E_{core} + \sum_{pq}^{active} h_{pq} \, \hat{a}^\dagger_p \hat{a}_q
    + \frac{1}{2}\sum_{pqrs}^{active} h_{pqrs} \, \hat{a}^\dagger_p \hat{a}^\dagger_q \hat{a}_r \hat{a}_s,
\end{equation}
where $E_{core}$ is the energy contribution from the frozen orbitals, and the summations are restricted to the active orbital subset. This approach substantially reduces the number of qubits, one- and two-electron terms, while maintaining chemical accuracy for valence properties~\cite{Romero2019}.

\subsubsection{\texorpdfstring{$\mathbb{Z}_2$}{Z2} Symmetry Tapering}
Many fermionic Hamiltonians possess discrete symmetries that form an Abelian $\mathbb{Z}_2$ group~\cite{Bravyi2017, Gard2020}. Each independent generator corresponds to a conserved quantum number, such as particle-number parity, total spin-parity, or molecular point-group symmetry, and partitions the Hilbert space into distinct symmetry sectors.

The $\mathbb{Z}_2$ symmetry converter identifies these commuting symmetries from the qubit Hamiltonian and exploits them to remove redundant qubits. For each symmetry generator with a known eigenvalue $\pm 1$, one qubit can be eliminated, yielding a reduced Hamiltonian:
\begin{equation}
\hat{H}_{tapered} = \mathcal{T}(\hat{H}_{qubit}),
\end{equation}
where $\mathcal{T}$ embeds the symmetry constraints into the remaining qubits. This process preserves the physical spectrum while significantly reducing qubit count and circuit depth~\cite{Bravyi2017, Setia2020}. Formally, the tapering procedure identifies an Abelian subgroup $\mathcal{S} \subset \mathcal{P}_N$ of the $N$-qubit Pauli group:
\begin{equation}
    \mathcal{P}_N = \pm\{\hat{\mathbb{I}}_{2}, \hat{\sigma}_x, \hat{\sigma}_y, \hat{\sigma}_z\}^{\otimes N},
\end{equation} 
such that all $\hat{s} \in \mathcal{S}$ commute with $\hat{H}_{qubit}$. Each $\hat{s}_j$ defines a $\mathbb{Z}_2$ symmetry with eigenvalues $\pm 1$. Clifford transformations $\hat{U}_j$ map these generators to single-qubit Pauli operators:
\begin{equation}
\hat{U}_j \hat{s}_j \hat{U}_j^\dagger = \sigma_x^{(q_j)},
\end{equation}
allowing qubits $q_j$ to be replaced by their eigenvalues and removed from the simulation~\cite{Setia2020}. Beyond $\mathbb{Z}_2$ symmetries, additional reductions based on spatial geometry, like point-group symmetries, can be combined with tapering for further compression of the active Hilbert space~\cite{Bravyi2017, Seki2020, Setia2020}. 

Together, the FC approximation and $\mathbb{Z}_2$ tapering represent two complementary reduction techniques that, when combined with efficient FTQMs, enable scalable quantum algorithms for molecular simulations within the resource limits of NISQ and near-term quantum hardware~\cite{OMalley2016}. Building on these theoretical foundations, the following section outlines the quantitative methodology employed to evaluate these strategies and their associated resource trade-offs.

\section{Methodology} \label{sec3:methodology}
The overall methodology for performing resource estimation in VQE-based molecular simulations is divided into four sequential stages, as summarized below. Each stage transforms the molecular input into progressively more universal gate-based QC hardware-relevant representations, culminating in circuit-level resource metrics.

\begin{figure}[htbp]
\centering
\includegraphics[width=\linewidth]{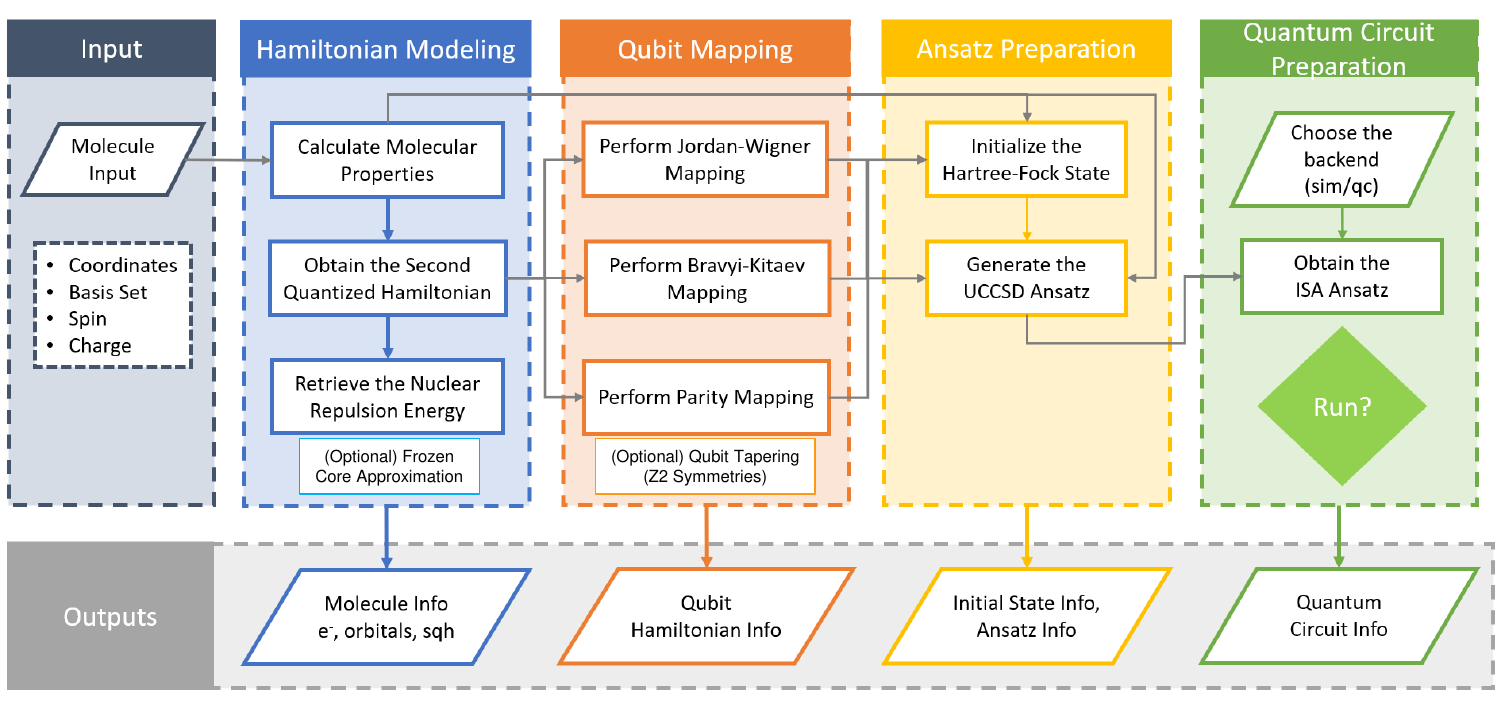}
\caption{Workflow for resource estimation in the VQE framework, showing the sequence from molecular input to Hamiltonian modeling, qubit mapping, ansatz construction, and circuit compilation, along with key outputs at each stage.}
\label{fig_2}
\end{figure}

The molecular geometry (in Cartesian coordinates), basis set, total spin, charge, and the desired backend are taken as input for resource estimation. The process is divided into four sequential stages: Hamiltonian Modeling, Qubit Mapping, Ansatz Preparation, and Quantum Circuit Preparation, each producing specific intermediate and final outputs, as shown in Figure~\ref{fig_2}.

\subsection{Hamiltonian Modeling}
In this stage, molecular properties are computed using the chosen basis set, and the second-quantized electronic Hamiltonian $\hat{H}_{sq}$ is obtained. The nuclear–nuclear repulsion energy term, $\hat{V}_{NN}$, is also retrieved. Hamiltonian reduction may be performed by enabling the frozen-core approximation, which decreases the number of active orbitals and thereby reduces both qubit requirements and gate counts. The outputs of this stage include the number of $\alpha$ and $\beta$ electrons, the number of spatial and spin orbitals, $\hat{V}_{NN}$, and the $\hat{H}_{sq}$.

\subsection{Qubit Mapping}
The $\hat{H}_{sq}$ is transformed into its qubit-representable form, $\hat{H}_{qubit}$, through FTQMs such as JW, BK, and Pa. To further reduce the Hamiltonian dimension, qubit tapering via $\mathbb{Z}_{2}$ symmetries can be applied, lowering the qubit count without introducing approximation. The outputs of this stage include the $\hat{H}_{qubit}$, the total number of Pauli strings, and their decomposition into $\hat{\sigma}_{x}$, $\hat{\sigma}_{y}$, $\hat{\sigma}_{z}$, and $\hat{\mathbb{I}}_{2}$ operators.

\subsection{Ansatz Preparation}
The Hartree-Fock~\cite{Fock1930} reference state $\ket{\psi_{HF}}$ is formulated using molecular parameters derived from the Hamiltonian modeling stage. Subsequently, the UCCSD ansatz is constructed to approximate the molecular ground state within the defined active space. The resulting outputs comprise detailed specifications of the initial state and the parametrized UCCSD ansatz corresponding to the target molecule.

\subsection{Quantum Circuit Preparation}
In the final stage, the UCCSD ansatz is compiled into the Instruction Set Architecture (ISA) format for a designated quantum simulator or hardware backend. This yields circuit-level resource information, including the number of qubits, total quantum gates, circuit depth, single-qubit and two-qubit gate counts, and individual gate breakdowns by type.

Together, these stages provide a comprehensive framework for assessing the computational resources required to simulate a given molecular system on a specific quantum simulator/hardware backend. The implementation was carried out in \texttt{Python (v3.11)}~\cite{Van1995, Python311} using \texttt{NumPy (v2.2.1)}~\cite{Harris2020}, \texttt{PySCF (v2.7.0)}~\cite{Sun2015, Sun2018, Sun2020}, \texttt{Qiskit (v1.3.1)}~\cite{Ali2024}, \texttt{Qiskit-Aer (v0.15.1)}~\cite{Ali2024, QiskitAer2024}, and \texttt{Qiskit-Nature (v0.7.2)}~\cite{QiskitNature2023}. The results obtained from this workflow for the representative set of molecular systems are discussed in the following section.

\section{Results and Discussion} \label{sec4:results}

We first present the results obtained from VQE(UCCSD)-based resource estimation performed using the workflow described in Section~\ref{sec3:methodology} on two representative molecular systems: Methane ($CH_4$) and Fluoromethane ($CH_3F$). Their 3-dimensional (3D) molecular geometries are shown in Figure~\ref{fig_3}, and the corresponding atomic coordinates are provided in Table~\ref{tab:SI_geometries} of the Supplementary Information. Both geometries were generated using the \texttt{Avogadro} molecular editor~\cite{Hanwell2012} using its built-in universal force field (UFF) auto-optimisation tool~\cite{Rappe1992}. These molecules are employed here as input geometries for the purpose of illustrating the resource-estimation workflow. 

For a direct comparison, all calculations employed the same STO-3G basis set and the JW FTQM, without applying any Hamiltonian reduction techniques mentioned in Section~\ref{sec2.6:hamiltonian_reduction}. The ISA-level ansatz circuits were transpiled using optimization level~3 and the \texttt{Qiskit-Aer} simulator backend in \texttt{Qiskit}. The simulator was employed with its default transpilation configuration, corresponding to an all-to-all qubit connectivity model without a hardware-imposed coupling map. As a result, no initial layout selection or routing procedures were required. Circuits were decomposed into the default transpilation target gate set of the \texttt{Qiskit-Aer} backend, consisting of single-qubit gates \{U$_2(\phi,\lambda)$, R$_Z(\theta)$, R$_Y(\theta)$, H, S$_{X}$ ($\sqrt{X}$), S$_{X}$dg ($\sqrt{X}^\dagger$), X\} and two-qubit gates $\{\mathrm{CX}$, \texttt{UnitaryGate}~\cite{UnitaryGateQiskit}\}. Under these conditions, the transpilation process is deterministic and yields identical compiled circuits across repeated runs for a given input circuit. The same compilation configuration was consistently applied to all subsequent calculations presented in this work.

Table~\ref{tab1:res_est} summarizes the results from the Hamiltonian Modeling and Qubit Mapping stages. The outputs include the chosen basis set, number of $\alpha$ and $\beta$ electrons, number of molecular spatial and spin orbitals ($N_{\mathrm{so}}$), and the $\hat{V}_{NN}$. Additionally, we report the total number of terms in $\hat{H}_{sq}$ ($N_{\mathrm{sqh}}$), the number of Pauli strings in the $\hat{H}_{qubit}$ ($N_{\mathrm{PS}}$), and the respective counts of $\hat{\sigma}_{x}$, $\hat{\sigma}_{y}$, $\hat{\sigma}_{z}$, and $\hat{\mathbb{I}}_{2}$ ($N_X$, $N_Y$, $N_Z$, $N_I$). These metrics collectively characterize the algebraic complexity and qubit-space dimensionality of each system. Subsequently, Table~\ref{tab2:res_est} presents the results from the Ansatz Preparation and Quantum Circuit Preparation stages. The outputs include the total number of qubits or circuit width ($W$), variational parameters ($P$), circuit depth ($D$), and total quantum gates ($G$), followed by a breakdown of the quantum gates into single-qubit, two-qubit, and arbitrary gate categories. These results characterize the relevant hardware-agnostic quantum circuit-level resource footprint for executing VQE(UCCSD).

\begin{figure}[htbp]
\centering
\includegraphics[width=\linewidth]{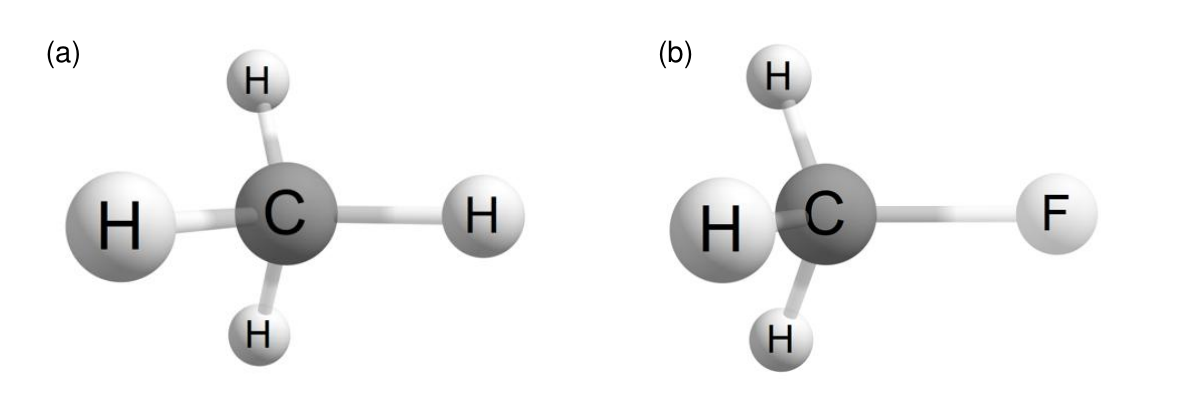}
\caption{3D molecular structures: (a) Methane ($CH_{4}$) and (b) Fluoromethane ($CH_{3}F$).}
\label{fig_3}
\end{figure}

\begin{table}[htbp]
\centering
\caption{\label{tab1:res_est}
Resource estimation results for $CH_{4}$ and $CH_{3}F$. Section~1 of the table presents the molecular information, followed by the $\hat{H}_{sq}$ information in Section~2, and finally the $\hat{H}_{qubit}$ information in Section~3. In Section~1, $\hat{V}_{NN}$ denotes the nuclear–nuclear repulsion energy, as given in Eq.~(\ref{eq:electronic_hamiltonian}); $Ha$ denotes Hartree, and $e^-$ represents an electron. In Section~2, the Number of Terms refers to the number of one- and two-body operators in $\hat{H}_{sq}$, as shown in Eq.~(\ref{eq:second_quantized_hamiltonian}). In Section~3, the Number of Pauli Strings $N_{\mathrm{PS}}$ denotes the unique Pauli strings that compose $\hat{H}_{qubit}$, and the counts of $\hat{\sigma}_{x}$, $\hat{\sigma}_{y}$, and $\hat{\sigma}_{z}$ ($N_{X}$, $N_{Y}$, $N_{Z}$, $N_{I}$) indicate the total number of the corresponding Pauli operators appearing across all Pauli strings in $\hat{H}_{qubit}$, as given by Eq.~(\ref{eq:qubit_hamiltonian}).}

\begin{tabular}{|l|r|r|}
\hline
\textbf{Property} & \textbf{Methane ($CH_{4}$)} & \textbf{Fluoromethane ($CH_{3}F$)} \\
\hline
\multicolumn{3}{|c|}{\textbf{1. Molecular Information}} \\
\hline
Basis-Set & STO-3G & STO-3G \\
\hline
Number of $\alpha$ electrons ($e^-$) & 5 & 9 \\
\hline
Number of $\beta$ electrons ($e^-$) & 5 & 9 \\
\hline
Number of Spatial Orbitals & 9 & 13 \\
\hline
Number of Spin Orbitals ($N_{\mathrm{so}}$) & 18 & 26 \\
\hline
$\hat{V}_{NN}$ in $Ha$ & 13.408333940368452 & 37.83061899847712 \\
\hline
\multicolumn{3}{|c|}{\textbf{2. Second-Quantized Hamiltonian $\hat{H}_{sq}$ Information}} \\
\hline
Number of Terms ($N_{\mathrm{sqh}}$) & 25,350 & 98,002 \\
\hline
\multicolumn{3}{|c|}{\textbf{3. Qubit Hamiltonian $\hat{H}_{qubit}$ Information}} \\
\hline
Number of Pauli Strings ($N_{\mathrm{PS}}$) & 8,172 & 28,984 \\
\hline
Number of $\hat{\sigma}_{x}$ [\%] ($N_{X}$) & 13,744 [09.34\%] & 50,648 [06.72\%] \\
\hline
Number of $\hat{\sigma}_{y}$ [\%] ($N_{Y}$) & 13,744 [09.34\%] & 50,648 [06.72\%] \\
\hline
Number of $\hat{\sigma}_{z}$ [\%]($N_{Z}$) & 30,612 [20.81\%] & 166,260 [22.06\%] \\
\hline
Number of $\hat{\mathbb{I}}_{2}$ [\%] ($N_{I}$) & 88,996 [60.50\%] & 486,028 (64.50\%] \\
\hline
\end{tabular}
\end{table}

\begin{table}[htbp]
\centering
\caption{\label{tab2:res_est}
Resource estimation results for $CH_4$ and $CH_3F$. Section~1 of the table provides the ansatz information, while Section~2 presents the quantum circuit information, followed by Sections~2.1, 2.2, and 2.3, which give detailed information on the corresponding single-qubit, two-qubit, and arbitrary gates, respectively. The quantum gate decomposition shown in Section~2 corresponds to using \texttt{Qiskit-Aer} as the backend with transpilation optimization level~3 in \texttt{Qiskit}. Additional quantum gate information can be found in the Qiskit documentation~\cite{circuitlib2025}.}
\begin{tabular}{|l|r|r|}
\hline
\textbf{Property / Gate} & \textbf{Methane ($CH_4$)} & \textbf{Fluoromethane ($CH_3F$)} \\
\hline
\multicolumn{3}{|c|}{\textbf{1. Ansatz Information}} \\
\hline
No. of Qubits ($W$) & 18 & 26 \\
\hline
No. of Parameters ($P$) & 560 & 1,800 \\
\hline
\multicolumn{3}{|c|}{\textbf{2. Quantum Circuit Information}} \\
\hline
Circuit Depth ($D$) & 70,182 & 3,12,660 \\
\hline
Total Gates ($G$) & 77,787 & 3,37,423 \\
\hline
Single-Qubit Gates & 12,673 & 41,985 \\
\hline
Two-Qubit Gates & 62,468 & 2,86,892 \\
\hline
Arbitrary Gates & 2,646 & 8,546 \\
\hline
\multicolumn{3}{|c|}{\textbf{2.1. Single-Qubit Gates Information}} \\
\hline
U$_2(\phi, \lambda)$ & 7,339 & 24,475 \\
\hline
R$_Z(\theta)$ & 4,240 & 13,968 \\
\hline
H & 529 & 1,733 \\
\hline
S$_X$ or $\sqrt{X}$ & 462 & 1,518 \\
\hline
S$_X$dg or $\sqrt{X}^{\dagger}$& 95 & 275 \\
\hline
X & 8 & 16 \\
\hline
\multicolumn{3}{|c|}{\textbf{2.2. Two-Qubit Gates Information}} \\
\hline
CX / CNOT & 62,468 & 2,86,892 \\
\hline
\multicolumn{3}{|c|}{\textbf{2.3. Arbitrary-Qubit Gates Information}} \\
\hline
Unitary & 2,646 & 8,546 \\
\hline
\end{tabular}
\end{table}

To further evaluate scalability, we extended our analysis across a series of molecules of increasing electronic and structural complexity, including $H_{2}$, $LiH$, $HF$, $BeH_2$, $H_2O$, $N_{2}$, $O_{2}$, $CO$, $NH_{3}$, $CH_{4}$, $C_{2}H_{2}$, $H_{2}O_{2}$, and $C_{2}H_{4}$, in the STO-3G basis set. For each molecule, the input geometry was obtained by performing a full geometry optimisation at the CCSD/STO-3G level of theory using \texttt{PySCF} quantum chemistry package~\cite{Sun2020}. All optimised atomic coordinates are tabulated in Table~\ref{tab:SI_geometries_CCSD} of the Supplementary Information. For each molecule, resource estimation was performed under four configurations combining Hamiltonian reduction options: frozen-core approximation (F) and $\mathbb{Z}_{2}$ symmetries, qubit tapering (T), with boolean combinations of:
\begin{equation*}
    (\mathrm{F}, \mathrm{T}) \rightarrow (False, False), (True, False), (False,
    True), (True, True).
\end{equation*}
Each configuration was evaluated using the JW, BK, and Pa FTQMs, and all circuits were transpiled at optimization level~3 for the \texttt{Qiskit-Aer} simulator backend under the same compilation settings described above. The corresponding Hamiltonian and circuit-level resource counts for all configurations are reported in Tables~\ref{tab:raw_data_A} and~\ref{tab:raw_data_B}.

In the trivial case of $H_{2}$, where no core orbitals are available for freezing, $\mathbb{Z}_{2}$ tapering effectively reduces the original four-qubit $\hat{H}_{qubit}$ to a single-qubit representation, illustrating the maximum achievable simplification. A structurally informative pattern emerges when examining the width reduction for the systems under study, as shown in Table~\ref{tab:width_reduction}. The underlying raw values from which these reduction factors are computed are listed in Table~\ref{tab:raw_data_B}, with the corresponding reduction ratios provided in Table~\ref{tab:reductions_data}. The JW and BK mappers yield identical qubit reductions across all molecules and all configurations. This equivalence arises as both encodings preserve the full $N$-qubit active space before any symmetry reduction, meaning the number of qubits that can be eliminated through tapering or the frozen-core approximation is governed entirely by the molecular symmetry content rather than the choice of mapping. The Pa mapper, in contrast, provides a default two-qubit overhead saving. This reduction stems from two independent $\mathbb{Z}_{2}$ symmetries, whose fixed eigenvalues allow two qubits to be removed at the encoding level, as detailed in Section~\ref{sec2.5:ftqm}. However, because these symmetries are already incorporated into the encoding, the number of additional exploitable $\mathbb{Z}_{2}$ symmetries available for post-mapping tapering is reduced or eliminated~\cite{Bravyi2017}. Specifically, for $NH_3$, $CH_4$, $C_2H_2$, and $H_2O_2$, the Pa mapper under the tapering-only (T) configuration yields no further width reduction, as confirmed by the $0.0\%$ entries in Table~\ref{tab:width_reduction}. We note that this interaction between the intrinsic symmetry capacity of the mapper and the molecule's symmetry content constitutes an important constraint that practitioners should consider when selecting a mapping strategy.

\begin{table}[htbp]
\centering
\setlength{\tabcolsep}{5pt}
\caption{Circuit width reduction (in \%) relative to the no-tapering, no-frozen-core baseline (-) for each molecule, qubit mapper, and reduction strategy combination. Reduction percentage is computed as $100 \times (Q_{-} - Q_{\mathrm{combo}})/Q_{-}$, where $Q$ denotes the total qubit count. A value of $x\%$ indicates that the optimised circuit requires $x\%$ fewer qubits than the baseline. T\,=\,tapering only; F\,=\,frozen core only; TF\,=\,both tapering and frozen core applied. Summary statistics (min, max, mean, median) are computed across all 12 molecules per column.}
\label{tab:width_reduction}

\begin{tabular}{|l|
c|c|c|
c|c|c|
c|c|c|}
\hline
& \multicolumn{3}{c|}{\textbf{Jordan--Wigner}} 
& \multicolumn{3}{c|}{\textbf{Bravyi--Kitaev}} 
& \multicolumn{3}{c|}{\textbf{Parity}} \\
\hline
\textbf{Molecule} & \textbf{T} & \textbf{F} & \textbf{TF} & \textbf{T} & \textbf{F} & \textbf{TF} & \textbf{T} & \textbf{F} & \textbf{TF} \\
\hline

\ce{LiH} & 33.3\% & 16.7\% & 50.0\% & 33.3\% & 16.7\% & 50.0\% & 20.0\% & 20.0\% & 40.0\% \\
\ce{HF} & 33.3\% & 16.7\% & 50.0\% & 33.3\% & 16.7\% & 50.0\% & 20.0\% & 20.0\% & 40.0\% \\
\ce{BeH2} & 35.7\% & 14.3\% & 50.0\% & 35.7\% & 14.3\% & 50.0\% & 25.0\% & 16.7\% & 41.7\% \\
\ce{H2O} & 21.4\% & 14.3\% & 35.7\% & 21.4\% & 14.3\% & 35.7\% & 8.3\% & 16.7\% & 25.0\% \\
\hline

\ce{NH3} & 12.5\% & 12.5\% & 25.0\% & 12.5\% & 12.5\% & 25.0\% & 0.0\% & 14.3\% & 14.3\% \\
\ce{CH4} & 11.1\% & 11.1\% & 22.2\% & 11.1\% & 11.1\% & 22.2\% & 0.0\% & 12.5\% & 12.5\% \\
\ce{O2} & 25.0\% & 20.0\% & 45.0\% & 25.0\% & 20.0\% & 45.0\% & 16.7\% & 22.2\% & 38.9\% \\
\ce{N2} & 20.0\% & 20.0\% & 40.0\% & 20.0\% & 20.0\% & 40.0\% & 11.1\% & 22.2\% & 33.3\% \\
\hline

\ce{CO} & 15.0\% & 20.0\% & 35.0\% & 15.0\% & 20.0\% & 35.0\% & 5.6\% & 22.2\% & 27.8\% \\
\ce{C2H2} & 8.3\% & 16.7\% & 25.0\% & 8.3\% & 16.7\% & 25.0\% & 0.0\% & 18.2\% & 18.2\% \\
\ce{H2O2} & 8.3\% & 16.7\% & 25.0\% & 8.3\% & 16.7\% & 25.0\% & 0.0\% & 18.2\% & 18.2\% \\
\ce{C2H4} & 10.7\% & 14.3\% & 25.0\% & 10.7\% & 14.3\% & 25.0\% & 3.8\% & 15.4\% & 19.2\% \\
\hline
\hline

\textbf{Min} & 8.3\% & 11.1\% & 22.2\% & 8.3\% & 11.1\% & 22.2\% & 0.0\% & 12.5\% & 12.5\% \\
\hline
\textbf{Max} & 35.7\% & 20.0\% & 50.0\% & 35.7\% & 20.0\% & 50.0\% & 25.0\% & 22.2\% & 41.7\% \\
\hline
\hline

\textbf{Mean} & 19.6\% & 16.1\% & 35.7\% & 19.6\% & 16.1\% & 35.7\% & 9.2\% & 18.2\% & 27.4\% \\
\hline
\textbf{Median} & 17.5\% & 16.7\% & 35.4\% & 17.5\% & 16.7\% & 35.4\% & 7.0\% & 18.2\% & 26.4\% \\
\hline

\end{tabular}
\end{table}

\begin{figure}[htbp]
    \centering
    \includegraphics[width=0.80\linewidth]{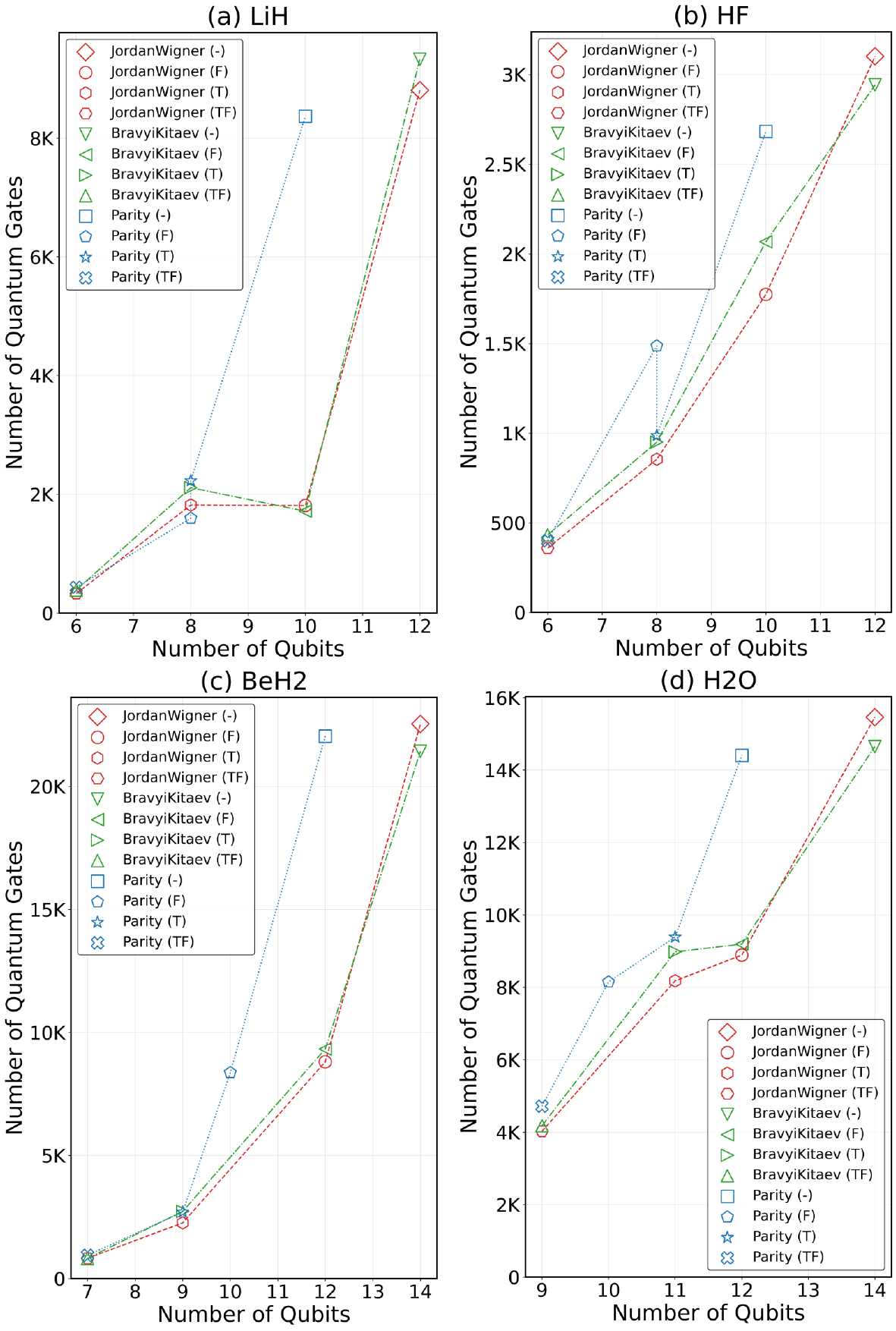}
    \caption{Scaling behaviour of qubit and gate requirements across molecular systems using JW, BK, Pa mappings and Hamiltonian reduction configurations.
    (a)-(d) correspond to the first set of molecular systems $(LiH, HF, BeH_2, \text{and } H_2O)$. Here, T denotes qubit tapering via $\mathbb{Z}_2$ symmetries, and F denotes the FC approximation. Symbols indicate the applied configurations: (–) both T and F are false, (T) only tapering is applied, (F) only frozen-core is applied, and (TF) both techniques are applied. Error bars are omitted as the resource-estimation procedure is deterministic.}
    \label{fig:4a}
\end{figure}

\begin{figure}[htbp]
    \centering
    \addtocounter{figure}{-1}
    \includegraphics[width=0.80\linewidth]{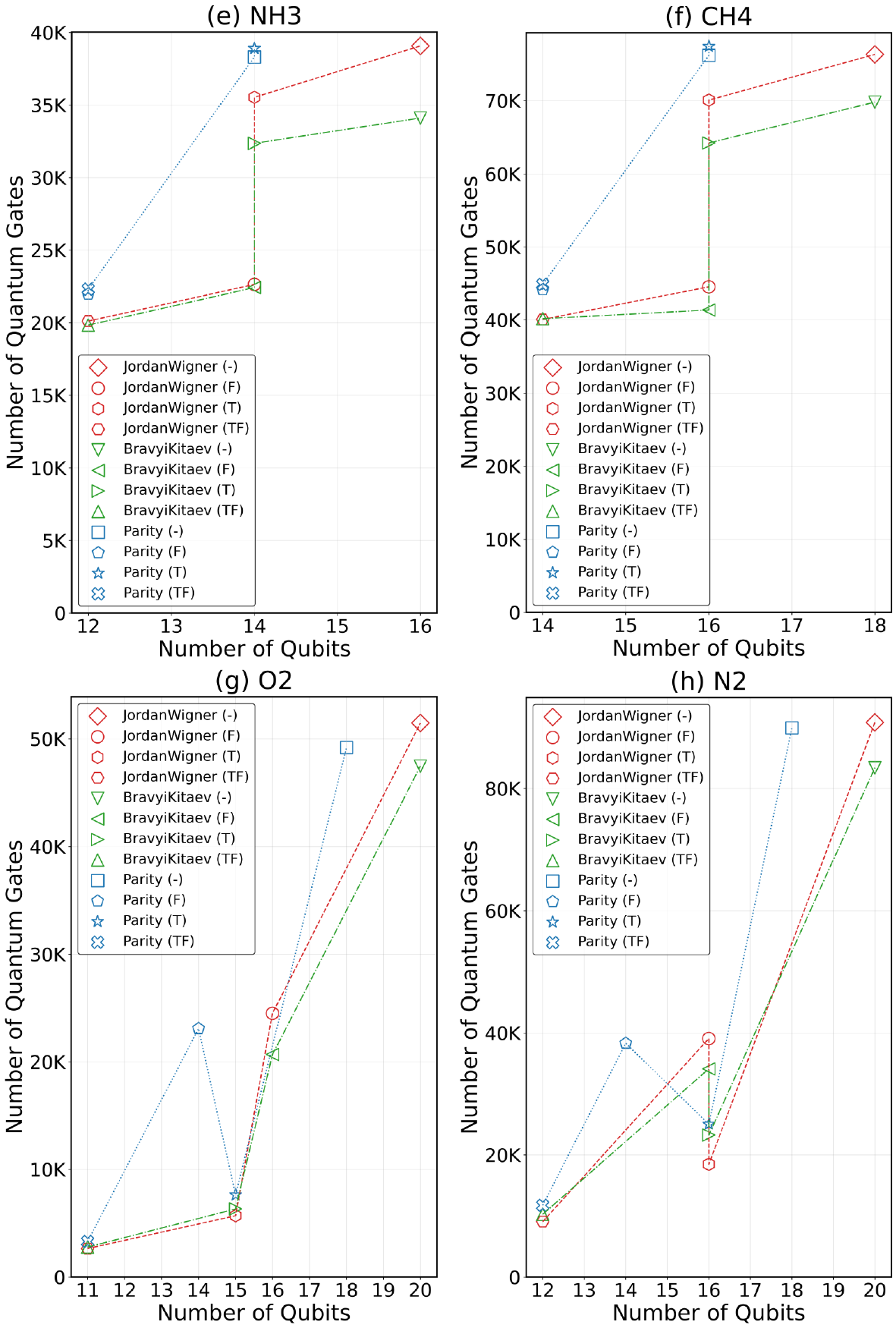}
    \caption{(continued) Scaling behaviour of qubit and gate requirements across molecular systems using JW, BK, Pa mappings and Hamiltonian reduction configurations.
    (e)-(h) correspond to the next set of molecular systems $(NH_3, CH_4, O_2,\text{and } N_2)$. Here, T denotes qubit tapering via $\mathbb{Z}_2$ symmetries, and F denotes the FC approximation. Symbols indicate the applied configurations: (–) both T and F are false, (T) only tapering is applied, (F) only frozen-core is applied, and (TF) both techniques are applied. Error bars are omitted as the resource-estimation procedure is deterministic.}
    \label{fig:4b}
\end{figure}

\begin{figure}[htbp]
    \centering
    \addtocounter{figure}{-1}
    \includegraphics[width=0.80\linewidth]{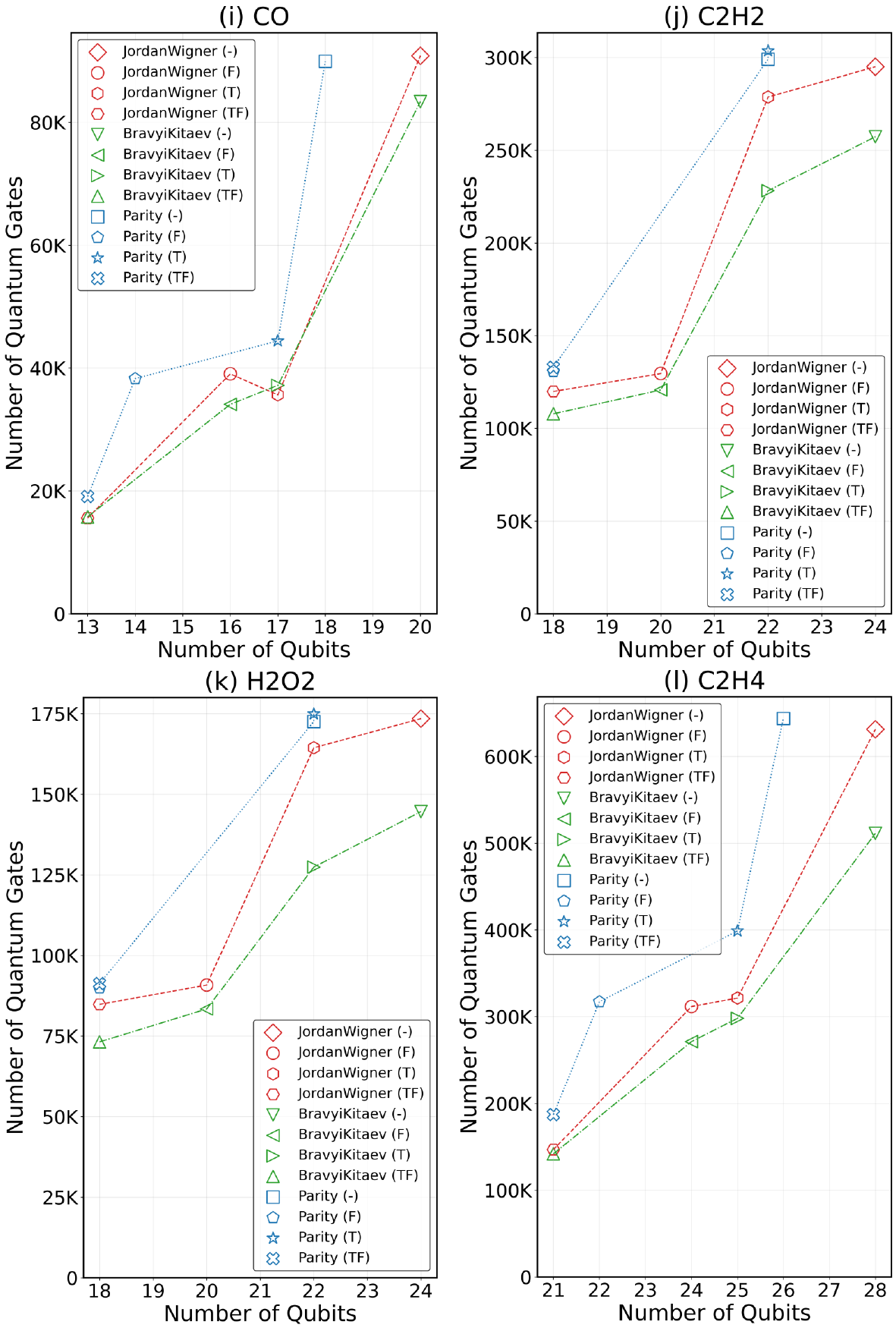}
    \caption{(continued) Scaling behaviour of qubit and gate requirements across molecular systems using JW, BK, Pa mappings and Hamiltonian reduction configurations.
    (i)-(l) correspond to the remaining molecular systems $(CO, C_2H_2, H_2O_2,\text{and } C_2H_4)$. Here, T denotes qubit tapering via $\mathbb{Z}_2$ symmetries, and F denotes the FC approximation. Symbols indicate the applied configurations: (–) both T and F are false, (T) only tapering is applied, (F) only frozen-core is applied, and (TF) both techniques are applied. Error bars are omitted as the resource-estimation procedure is deterministic.}
    \label{fig:4c}
\end{figure}

Beyond $H_2$, for the given set of molecules, tapering and frozen-core techniques yield progressively smaller reductions but still significantly lower the total gate count and qubit width, as shown in Figure~\ref{fig:4a}. A finer-grained inspection of the qubit widths, reported in Table~\ref{tab:width_reduction}, reveals that the frozen-core approximation (F) delivers more uniform width reductions across molecules. This reduction ranges from $11.1\%$ to $20.0\%$ for JW and BK, whereas tapering alone (T) exhibits considerably higher variance, spanning from $0\%$ to $35.7\%$ depending on the available $\mathbb{Z}_2$ symmetries of the system. This uniformity is physically intuitive: freezing core orbitals removes a fixed number of spin-orbitals proportional to the number of inner-shell electrons, which scales predictably with atomic composition, whereas tapering is highly sensitive to point-group symmetry and may yield no benefit when the applicable symmetry sectors are already saturated. When both techniques are combined (TF), the reductions compound, with the mean width reduction rising to $35.7\%$ for JW and BK and $27.4\%$ for the Pa mapper, as summarized in the statistical rows of Table~\ref{tab:width_reduction}. 

Complementing the qubit-width analysis, it is informative to examine how the Hamiltonian-level operator count, specifically the number of unique Pauli strings $N_{\text{PS}}$ in $\hat{H}_{\text{qubit}}$, responds to the same reduction strategies. This quantity is directly relevant to the measurement cost of VQE, since each unique Pauli string in $\hat{H}_{\text{qubit}}$ in principle requires a separate measurement circuit when evaluating expectation values, and the total number of measurements scales with both $N_{\text{PS}}$ and the statistical precision required~\cite{Wecker2015, Gonthier2022}. The corresponding reduction factors for $N_{\text{PS}}$ across all molecules, mappers, and configurations are summarised in Table~\ref{tab:NPS_reduction}. A structurally significant observation is that $N_{\text{PS}}$ reduction factors are identical across all three mappers for every molecule and configuration. This mapper independence arises as the count of unique Pauli strings is determined by the molecular symmetry and active-space structure rather than the specific FTQM, which affects the form of individual strings but not their total number. Consequently, it may be inferred that the choice of mapper has no bearing on measurement cost as quantified by $N_{\text{PS}}$.

In terms of the magnitude of reduction, the frozen-core approximation (F) and its combination with symmetry tapering (TF) deliver the most meaningful $N_{\text{PS}}$ compression, with mean reduction factors of $2.09\times$ and $2.21\times$, respectively, and peak values reaching $2.71\times$ for $O_2$ (F) and $2.75\times$ for $O_2$ (TF). By contrast, tapering alone (T) yields negligible $N_{\text{PS}}$ reduction for most molecules, with a mean of $1.03\times$ and a median of $1.00\times$, confirming that $\mathbb{Z}_2$ symmetry tapering primarily reduces the number of qubits and simplifies operator weights rather than eliminating large numbers of unique measurement bases. The notable exception is $H_2$, where the absence of core orbitals precludes the frozen-core approximation entirely, and the high point-group symmetry of the system allows tapering alone to reduce $N_{\text{PS}}$ from 15 to 3, a $5\times$ compression that reflects the maximum symmetry exploitation achievable for this well-studied two-electron system~\cite{Peruzzo2014, OMalley2016}; $H_2$ is accordingly excluded from the 12-molecule summary statistics reported here. For molecules such as $NH_3$, $CH_4$, $O_2$, $N_2$, $CO$, $C_2H_2$, $H_2O_2$, and $C_2H_4$, the tapering-only (T) reduction factor is exactly $1.00\times$, meaning that $N_{\mathrm{PS}}$ is entirely unchanged by symmetry tapering alone. These findings indicate that the frozen-core approximation is the dominant mechanism for reducing measurement overhead among the strategies examined here, and that its effect compounds only modestly when combined with tapering. It should be noted that this study focuses on circuit-level resource requirements, and a full quantitative analysis of measurement cost, including shot-count estimation and Pauli grouping strategies, is outside the scope of the present work and is deferred as a natural direction for future investigation.

\begin{table}[htbp]
\centering
\setlength{\tabcolsep}{5pt}
\caption{Pauli string count ($N_{\text{PS}}$) reduction factor  relative to the no-tapering, no-frozen-core baseline (-) for each molecule, qubit mapper, and reduction strategy combination. Reduction factor is computed as $N_{\text{PS},-}/N_{\text{PS,combo}}$, where $N_{\text{PS}}$ denotes the number of unique Pauli strings in the qubit Hamiltonian $\hat{H}_{\text{qubit}}$. A value of $k\times$ indicates the optimised Hamiltonian contains $k$ times fewer Pauli strings than the baseline. T\,=\,tapering only; F\,=\,frozen core only; TF\,=\,both tapering and frozen core applied. Summary statistics (min, max, mean, median) are computed across all 12 molecules per column.}
\label{tab:NPS_reduction}
\begin{tabular}{|l|
c|c|c|
c|c|c|
c|c|c|}
\hline
& \multicolumn{3}{c|}{\textbf{Jordan--Wigner}}
& \multicolumn{3}{c|}{\textbf{Bravyi--Kitaev}}
& \multicolumn{3}{c|}{\textbf{Parity}} \\
\hline
\textbf{Molecule} & \textbf{T} & \textbf{F} & \textbf{TF} & \textbf{T} & \textbf{F} & \textbf{TF} & \textbf{T} & \textbf{F} & \textbf{TF} \\
\hline
\ce{LiH}  & 1.13$\times$ & 2.29$\times$ & 2.73$\times$ & 1.13$\times$ & 2.29$\times$ & 2.73$\times$ & 1.13$\times$ & 2.29$\times$ & 2.73$\times$ \\
\ce{HF}   & 1.13$\times$ & 2.29$\times$ & 2.73$\times$ & 1.13$\times$ & 2.29$\times$ & 2.73$\times$ & 1.13$\times$ & 2.29$\times$ & 2.73$\times$ \\
\ce{BeH2} & 1.12$\times$ & 2.04$\times$ & 2.49$\times$ & 1.12$\times$ & 2.04$\times$ & 2.49$\times$ & 1.12$\times$ & 2.04$\times$ & 2.49$\times$ \\
\ce{H2O}  & 1.04$\times$ & 1.97$\times$ & 2.08$\times$ & 1.04$\times$ & 1.97$\times$ & 2.08$\times$ & 1.04$\times$ & 1.97$\times$ & 2.08$\times$ \\
\hline
\ce{NH3}  & 1.00$\times$ & 1.55$\times$ & 1.55$\times$ & 1.00$\times$ & 1.55$\times$ & 1.55$\times$ & 1.00$\times$ & 1.55$\times$ & 1.55$\times$ \\
\ce{CH4}  & 1.00$\times$ & 1.47$\times$ & 1.47$\times$ & 1.00$\times$ & 1.47$\times$ & 1.47$\times$ & 1.00$\times$ & 1.47$\times$ & 1.47$\times$ \\
\ce{O2}   & 1.00$\times$ & 2.71$\times$ & 2.75$\times$ & 1.00$\times$ & 2.71$\times$ & 2.75$\times$ & 1.00$\times$ & 2.71$\times$ & 2.75$\times$ \\
\ce{N2}   & 1.00$\times$ & 2.51$\times$ & 2.51$\times$ & 1.00$\times$ & 2.51$\times$ & 2.51$\times$ & 1.00$\times$ & 2.51$\times$ & 2.51$\times$ \\
\hline
\ce{CO}   & 1.00$\times$ & 2.50$\times$ & 2.50$\times$ & 1.00$\times$ & 2.50$\times$ & 2.50$\times$ & 1.00$\times$ & 2.50$\times$ & 2.50$\times$ \\
\ce{C2H2} & 1.00$\times$ & 2.02$\times$ & 2.02$\times$ & 1.00$\times$ & 2.02$\times$ & 2.02$\times$ & 1.00$\times$ & 2.02$\times$ & 2.02$\times$ \\
\ce{H2O2} & 1.00$\times$ & 1.93$\times$ & 1.93$\times$ & 1.00$\times$ & 1.93$\times$ & 1.93$\times$ & 1.00$\times$ & 1.93$\times$ & 1.93$\times$ \\
\ce{C2H4} & 1.00$\times$ & 1.76$\times$ & 1.77$\times$ & 1.00$\times$ & 1.76$\times$ & 1.77$\times$ & 1.00$\times$ & 1.76$\times$ & 1.77$\times$ \\
\hline
\hline
\textbf{Min}    & 1.00$\times$ & 1.47$\times$ & 1.47$\times$ & 1.00$\times$ & 1.47$\times$ & 1.47$\times$ & 1.00$\times$ & 1.47$\times$ & 1.47$\times$ \\
\hline
\textbf{Max}    & 1.13$\times$ & 2.71$\times$ & 2.75$\times$ & 1.13$\times$ & 2.71$\times$ & 2.75$\times$ & 1.13$\times$ & 2.71$\times$ & 2.75$\times$ \\
\hline
\hline
\textbf{Mean}   & 1.03$\times$ & 2.09$\times$ & 2.21$\times$ & 1.03$\times$ & 2.09$\times$ & 2.21$\times$ & 1.03$\times$ & 2.09$\times$ & 2.21$\times$ \\
\hline
\textbf{Median} & 1.00$\times$ & 2.03$\times$ & 2.29$\times$ & 1.00$\times$ & 2.03$\times$ & 2.29$\times$ & 1.00$\times$ & 2.03$\times$ & 2.29$\times$ \\
\hline
\end{tabular}
\end{table}

Turning to the number of quantum gates, the impact on total gate counts is more heterogeneous across the molecular set. As detailed in Table~\ref{tab:gates_reduction}, when both symmetry tapering and the frozen-core approximation are applied, the JW mapping exhibits reductions ranging from $1.91\times$ for $CH_4$ to $27.48\times$ for $BeH_2$, with a mean of $9.55\times$ and a median of $5.06\times$. The pronounced separation between the mean and median indicates that a small subset of molecules, principally $BeH_2$, $LiH$, and $O_2$, benefit disproportionately from favourable symmetry structure and a comparatively large frozen-core contribution. The complete set of reduction factors for all resource metrics is provided in Table~\ref{tab:reductions_data}. For the majority of molecules in the benchmark set, the gate reduction clusters more modestly around $2\times$ to $5\times$ under the combined application of both techniques. It should be noted that this variability has practical implications, as aggregate resource savings inferred from a limited number of highly symmetric molecular systems may overestimate the reductions achievable for a representative workload. Therefore, careful, system-specific resource estimation is warranted.

\begin{table}[htbp]
\centering
\setlength{\tabcolsep}{5pt}
\caption{Total gate count reduction factor ($N_{\times}$) relative to the no-tapering, no-frozen-core baseline (-) for each molecule, qubit mapper, and reduction strategy combination. Reduction factor is computed as $G_{-}/G_{\mathrm{combo}}$, where $G$ denotes the total gate count. A value of $k\times$ indicates the optimised circuit requires $k$ times fewer gates than the baseline. Values below $1.00\times$ indicate a marginal gate count increase versus baseline. T\,=\,tapering only; F\,=\,frozen core only; TF\,=\,both tapering and frozen core applied. Summary statistics (min, max, mean, median) are computed across all 12 molecules per column.}
\label{tab:gates_reduction}

\begin{tabular}{|l|
c|c|c|
c|c|c|
c|c|c|}
\hline
& \multicolumn{3}{c|}{\textbf{Jordan--Wigner}} 
& \multicolumn{3}{c|}{\textbf{Bravyi--Kitaev}} 
& \multicolumn{3}{c|}{\textbf{Parity}} \\
\hline
\textbf{Molecule} & \textbf{T} & \textbf{F} & \textbf{TF} & \textbf{T} & \textbf{F} & \textbf{TF} & \textbf{T} & \textbf{F} & \textbf{TF} \\
\hline

\ce{LiH} & 4.84$\times$ & 4.86$\times$ & 26.69$\times$ & 4.42$\times$ & 5.44$\times$ & 24.05$\times$ & 3.75$\times$ & 5.24$\times$ & 19.24$\times$ \\
\ce{HF} & 3.63$\times$ & 1.75$\times$ & 8.66$\times$ & 3.10$\times$ & 1.42$\times$ & 6.81$\times$ & 2.72$\times$ & 1.80$\times$ & 6.71$\times$ \\
\ce{BeH2} & 9.96$\times$ & 2.56$\times$ & 27.48$\times$ & 7.85$\times$ & 2.30$\times$ & 26.17$\times$ & 8.15$\times$ & 2.63$\times$ & 23.42$\times$ \\
\ce{H2O} & 1.89$\times$ & 1.74$\times$ & 3.84$\times$ & 1.63$\times$ & 1.59$\times$ & 3.50$\times$ & 1.53$\times$ & 1.77$\times$ & 3.05$\times$ \\
\hline

\ce{NH3} & 1.10$\times$ & 1.73$\times$ & 1.94$\times$ & 1.05$\times$ & 1.52$\times$ & 1.72$\times$ & 0.98$\times$ & 1.74$\times$ & 1.72$\times$ \\
\ce{CH4} & 1.09$\times$ & 1.71$\times$ & 1.91$\times$ & 1.09$\times$ & 1.69$\times$ & 1.74$\times$ & 0.98$\times$ & 1.73$\times$ & 1.70$\times$ \\
\ce{O2} & 9.02$\times$ & 2.10$\times$ & 19.42$\times$ & 7.49$\times$ & 2.30$\times$ & 17.02$\times$ & 6.44$\times$ & 2.13$\times$ & 14.88$\times$ \\
\ce{N2} & 4.92$\times$ & 2.33$\times$ & 9.99$\times$ & 3.59$\times$ & 2.45$\times$ & 8.10$\times$ & 3.59$\times$ & 2.35$\times$ & 7.64$\times$ \\
\hline

\ce{CO} & 2.55$\times$ & 2.33$\times$ & 5.83$\times$ & 2.24$\times$ & 2.45$\times$ & 5.29$\times$ & 2.02$\times$ & 2.35$\times$ & 4.71$\times$ \\
\ce{C2H2} & 1.06$\times$ & 2.28$\times$ & 2.46$\times$ & 1.13$\times$ & 2.13$\times$ & 2.39$\times$ & 0.99$\times$ & 2.29$\times$ & 2.25$\times$ \\
\ce{H2O2} & 1.06$\times$ & 1.91$\times$ & 2.05$\times$ & 1.14$\times$ & 1.73$\times$ & 1.98$\times$ & 0.99$\times$ & 1.92$\times$ & 1.89$\times$ \\
\ce{C2H4} & 1.96$\times$ & 2.02$\times$ & 4.30$\times$ & 1.72$\times$ & 1.89$\times$ & 3.60$\times$ & 1.61$\times$ & 2.03$\times$ & 3.44$\times$ \\
\hline
\hline

\textbf{Min} & 1.06$\times$ & 1.71$\times$ & 1.91$\times$ & 1.05$\times$ & 1.42$\times$ & 1.72$\times$ & 0.98$\times$ & 1.73$\times$ & 1.70$\times$ \\
\hline
\textbf{Max} & 9.96$\times$ & 4.86$\times$ & 27.48$\times$ & 7.85$\times$ & 5.44$\times$ & 26.17$\times$ & 8.15$\times$ & 5.24$\times$ & 23.42$\times$ \\
\hline
\hline

\textbf{Mean} & 3.59$\times$ & 2.28$\times$ & 9.55$\times$ & 3.04$\times$ & 2.24$\times$ & 8.53$\times$ & 2.81$\times$ & 2.33$\times$ & 7.55$\times$ \\
\hline
\textbf{Median} & 2.25$\times$ & 2.06$\times$ & 5.06$\times$ & 1.98$\times$ & 2.01$\times$ & 4.45$\times$ & 1.81$\times$ & 2.08$\times$ & 4.08$\times$ \\
\hline

\end{tabular}
\end{table}

The results indicate that although the JW mapping provides the most straightforward implementation, the BK mapping can yield superior compactness in specific cases. For larger molecules such as $CO$, $C_2H_2$, $H_2O_2$, and $C_2H_4$, the BK mapping tends to produce shallower circuits with fewer total gates, as illustrated in Figure~\ref{fig:4c}. The relative positioning of these three mappers can be further clarified by examining their statistical summary, as presented in Table~\ref{tab:gates_reduction}. Across all molecules, the JW mapper achieves the highest mean gate reduction when both tapering and frozen-core approximation are applied ($9.55\times$), followed by BK ($8.53\times$) and Pa ($7.55\times$). The ordering is reversed and compressed when only the frozen-core approximation is applied, with all three mappers statistically near-equivalent (means of $2.28\times$, $2.24\times$, and $2.33\times$, respectively), suggesting that the frozen-core approximation alone is largely mapper-agnostic in its effect. The largest absolute spread between mappers emerges when only symmetry tapering is applied, where JW's mean of $3.59\times$ substantially exceeds Pa's $2.81\times$, reinforcing the observation that $\mathbb{Z}_2$ tapering interacts most favourably with the JW encoding.

A more consequential observation emerges for the Pa mapper under tapering-only configurations, where a marginal gate-count increase relative to the unoptimised baseline is observed for $NH_3$, $CH_4$, $C_2H_2$, and $H_2O_2$, with reduction factors of approximately $0.98\times$ to $0.99\times$, as shown in Table~\ref{tab:gates_reduction}. The full reduction data for these cases are reported in Table~\ref{tab:reductions_data}. This counterintuitive outcome arises because the parity encoding redistributes fermionic occupation information into cumulative parity operators, which can increase the entangling structure of the resulting Pauli strings. When this increase is not offset by a sufficient reduction in operator count from tapering, a net gate overhead may result. Importantly, the frozen-core approximation restores the benefit for these molecules: both the frozen-core approximation alone and its combination with symmetry tapering yield positive reductions across all molecules and all mappers, as confirmed by the absence of sub-unity entries in the corresponding columns of Table~\ref{tab:gates_reduction}. It may therefore be inferred that this finding provides a practical guideline: when employing the Pa mapper, the frozen-core approximation should be applied as a minimum prerequisite to ensure that symmetry-based optimisations do not inadvertently increase circuit costs.

Finally, it is worth noting that the relationship between qubit count and total gate count is not linear across the molecule set, as shown in Figure~\ref{fig:4a}. Molecules such as $BeH_2$ and $O_2$, despite having comparable qubit counts to $LiH$ and $HF$ after both symmetry tapering and frozen-core approximations are applied, exhibit markedly different absolute gate counts. This non-linearity reflects the sensitivity of gate complexity to operator density in the qubit Hamiltonian, a property governed by molecular symmetry and orbital structure.

These trends suggest a practical hierarchy to guide informed FTQM selection and compilation strategies for near-term molecular VQE-UCCSD experiments: for systems where only tapering is feasible, JW or BK should be favoured; where the frozen-core approximation is available, all three mappers are broadly competitive; and where both reductions are applicable, JW and BK deliver the greatest aggregate gate savings, with BK providing an additional advantage in total number of quantum gates for the larger molecules as noted previously in this study.

\section{Conclusion} \label{sec5:conclusion}
This work presents a systematic framework for hardware-agnostic quantum resource estimation in molecular simulations based on VQE using the UCCSD ansatz. By integrating Hamiltonian modeling, qubit mapping, ansatz construction, and circuit compilation within a unified workflow, we provide an end-to-end assessment of the general computational requirements for executing chemistry-relevant problems on NISQ and near-term quantum hardware, with the physical circuit-level metrics reported here serving as a reference baseline for future logical-resource analyses targeting fault-tolerant quantum systems. Our results demonstrate that FTQMs and Hamiltonian reduction strategies significantly affect quantum circuit complexity. Techniques such as frozen-core approximation and $\mathbb{Z}_{2}$ symmetry tapering can substantially lower qubit counts and quantum gate operations without sacrificing physical accuracy. The scaling analysis across the benchmarked molecular systems quantitatively establishes that optimal combinations of mapping and reduction can reduce total qubits by up to $\approx 50\%$, Pauli string counts by up to $\approx 2.75 \times$ and quantum gates by up to $\approx 27.5\times$, compared to the unreduced Hamiltonian under identical active-space and compilation settings, thereby improving the feasibility of molecular simulations on NISQ and near-term quantum hardware.

A natural extension of this work is the systematic integration of other FTQMs, particularly those based on ternary-tree constructions, into the resource estimation pipeline. Recent advances, such as optimal tree mappings \cite{Jiang2020}, compact and hierarchical encodings \cite{Derby2021, Miller2023}, ultrafast hybrid mappings \cite{OBrien2024}, and physically inspired low-entanglement schemes \cite{Parella2024}, have demonstrated significant reductions in both qubit overhead and Pauli operator weight compared to canonical JW and BK mappings. Incorporating, benchmarking, and cross-comparing these mappings within our unified workflow would enable a more precise characterization of mapping-ansatz-hardware co-dependencies. In particular, adaptive and hardware-aware ternary-tree frameworks such as HATT \cite{Liu2025} and Clifford-optimized heuristic mappings \cite{Yu2025} offer structured pathways for minimizing circuit depth and two-qubit gate counts. Evaluating these emerging FTQMs across a wider set of chemically relevant systems, therefore, represents a key future direction toward achieving increasingly scalable and resource-efficient VQE and QPE implementations.

Beyond electronic-structure Hamiltonians, an important long-term extension of this framework is the incorporation of non-B.O. formulations such as NOMO, MCMO, NEO, and CNEO. Integrating these non-B.O. Hamiltonians into a unified resource-estimation workflow would enable quantitative assessments of the quantum resources required for simulations where nuclear quantum effects are non-negligible~\cite{Schiffer1994, Tuckerman1997, Morrone2008, Reece2009, Xin2011, Klinman2013, Wang2014, Ceriotti2016, Markland2018, Mengxu2024, Han2025, Ugur2025}. Developing non-B.O. Hamiltonians and ansätze for QPE and VQE, therefore, represents a promising direction for future research~\cite{Veis2016, Kovyrshin2023, Nykanen2023, Cabral2025}.

From an applied and industrial perspective, this study provides actionable insights for selecting mapping strategies, circuit optimizations, and molecular configurations compatible with specific backend constraints. From the perspective of practical VQE execution, a quantitative analysis of measurement overhead, encompassing shot-count estimation, Pauli grouping strategies, and covariance-based measurement reduction, represents an important complement to the circuit-level resource estimates presented here, and is identified as a direction for future work. The resource estimation workflow developed here can be readily extended for resource-aware benchmarking, algorithm-hardware co-design, and automated pipeline integration in quantum computational chemistry applications.

\section*{Acknowledgements}
 The authors would like to express their appreciation to the advisors at Qclairvoyance Quantum Labs for their support, constructive discussions, and inspiration throughout the preparation of this work.

\section*{Funding}
No funding was received for this research.

\section*{Conflict of Interest}
R.M. and R.V. are paid consultants at Qclairvoyance Quantum Labs. The other authors declare no competing interests.

\section*{Data Availability}
The data that support the findings of this study are available in the supplementary material of this article.

\bibliography{references}

\newpage
\setcounter{table}{0}
\renewcommand{\thetable}{S\arabic{table}}

\section*{Supplementary Information}

This Supplementary Information provides detailed structural and raw resource data supporting the analysis presented in Section~\ref{sec4:results} of the main manuscript. The contents are organised to ensure full transparency and reproducibility of the quantum resource estimation workflow.

First, we report the molecular geometries used throughout the study. For the small illustrative systems ($CH_4$ and $CH_3F$), force-field-optimised coordinates generated using the \texttt{Avogadro} molecular editor with the universal force field (UFF) are provided in Table~\ref{tab:SI_geometries}. For all 13 benchmark molecules used in the scalability analysis, coupled-cluster singles and doubles (CCSD) optimised geometries at the STO-3G level of theory, computed using \texttt{PySCF}, are listed in Table~\ref{tab:SI_geometries_CCSD}. All coordinates are reported in \AA\ and rounded to four decimal places. Atom ordering follows the corresponding \texttt{.xyz} file outputs to ensure direct reproducibility.

Second, we provide complete raw Hamiltonian- and circuit-level resource counts for every molecule, fermion-to-qubit mapping (Jordan--Wigner, Bravyi--Kitaev, and Parity), and reduction configuration (no tapering/no frozen core, tapering only, frozen core only, and both combined) in Tables~\ref{tab:raw_data_A} and~\ref{tab:raw_data_B}. Reported quantities include spin-orbital counts, Hamiltonian term counts, Pauli string statistics, individual Pauli operator counts, circuit width, circuit depth, total gate counts, and number of variational parameters.

Finally, we include explicit reduction factors computed relative to the no-tapering/no-frozen-core baseline for each molecule and mapping in Table~\ref{tab:reductions_data}. These tables provide the numerical basis for all percentage and multiplicative reductions discussed in the main text.

Together, these data ensure that all reported scaling trends, resource reductions, and comparative statements in the manuscript are fully traceable to their underlying raw quantities.


\begin{longtable}{llrrr}
\caption{Force-field-optimised atomic coordinates (\AA) for \ce{CH4} and \ce{CH3F}, as generated by the \texttt{Avogadro} molecular editor~\cite{Hanwell2012} using the built-in universal force field (UFF) auto-optimisation tool~\cite{Rappe1992}. These geometries serve as illustrative inputs for the resource-estimation workflow presented in Section~\ref{sec4:results} and are not intended as high-accuracy reference structures. Atom ordering follows the output of the \texttt{.xyz} file produced by Avogadro.}
\label{tab:SI_geometries}\\
\toprule
\textbf{Molecule} & \textbf{Atom} & \textbf{x (\AA)} & \textbf{y (\AA)} & \textbf{z (\AA)} \\
\midrule
\endfirsthead
\caption[]{(\textit{continued})}\\
\toprule
\textbf{Molecule} & \textbf{Atom} & \textbf{x (\AA)} & \textbf{y (\AA)} & \textbf{z (\AA)} \\
\midrule
\endhead
\midrule \multicolumn{5}{r}{\textit{Continued on next page}} \\
\endfoot
\bottomrule
\endlastfoot

\textit{Methane (\ce{CH4})}
  & C & $-4.9566$ & $\phantom{-}4.6276$ & $\phantom{-}0.00000$ \\
  & H & $-5.8816$ & $\phantom{-}4.7456$ & $-0.5688$           \\
  & H & $-4.9356$ & $\phantom{-}5.3539$ & $\phantom{-}0.8155$ \\
  & H & $-4.9085$ & $\phantom{-}3.6170$ & $\phantom{-}0.4113$ \\
  & H & $-4.1009$ & $\phantom{-}4.7941$ & $-0.6580$           \\
\midrule
\textit{Fluoromethane (\ce{CH3F})}
  & C & $-4.9578$ & $\phantom{-}4.5872$ & $-0.0454$           \\
  & H & $-5.8862$ & $\phantom{-}4.7153$ & $-0.6056$           \\
  & F & $-4.9316$ & $\phantom{-}5.4894$ & $\phantom{-}0.9676$ \\
  & H & $-4.9092$ & $\phantom{-}3.5822$ & $\phantom{-}0.3785$ \\
  & H & $-4.0983$ & $\phantom{-}4.7641$ & $-0.6951$           \\

\end{longtable}


\begin{longtable}{llrrr}
\caption{CCSD-optimised atomic coordinates (\AA) for all 13 molecules used in the scalability analysis of Section~\ref{sec4:results}. Geometry optimisations were performed at the CCSD/STO-3G level of theory using the \texttt{PySCF} quantum chemistry package~\cite{Sun2020}, with the \texttt{geometric} solver employed for the optimisation procedure. Coordinates are reported to four decimal places. Atom ordering follows the output of the \texttt{.xyz} files produced by\texttt{PySCF}.}
\label{tab:SI_geometries_CCSD}\\
\toprule
\textbf{Molecule} & \textbf{Atom} & \textbf{x (\AA)} & \textbf{y (\AA)} & \textbf{z (\AA)} \\
\midrule
\endfirsthead
\caption[]{(\textit{continued})}\\
\toprule
\textbf{Molecule} & \textbf{Atom} & \textbf{x (\AA)} & \textbf{y (\AA)} & \textbf{z (\AA)} \\
\midrule
\endhead
\midrule \multicolumn{5}{r}{\textit{Continued on next page}} \\
\endfoot
\bottomrule
\endlastfoot

\textit{Hydrogen (\ce{H2})}
  & H & $-8.0162$ & $\phantom{-}4.7102$ & $\phantom{-}0.0247$ \\
  & H & $-8.6385$ & $\phantom{-}4.7896$ & $-0.3580$ \\
\midrule

\textit{Lithium Hydride (\ce{LiH})}
  & Li & $-1.3219$ & $\phantom{-}3.0763$ & $\phantom{-}0.0338$ \\
  & H  & $-0.6203$ & $\phantom{-}2.1491$ & $\phantom{-}1.0550$ \\
\midrule

\textit{Hydrogen Fluoride (\ce{HF})}
  & F & $-9.3450$ & $\phantom{-}1.4356$ & $\phantom{-}0.0011$ \\
  & H & $-8.3880$ & $\phantom{-}1.6991$ & $-0.0613$ \\
\midrule

\textit{Beryllium Hydride (\ce{BeH2})}
  & Be & $-0.9837$ & $\phantom{-}3.4084$ & $\phantom{-}0.1009$ \\
  & H  & $-1.6817$ & $\phantom{-}2.9596$ & $-0.9204$ \\
\end{longtable}

\begin{landscape}
\begin{longtable}{llcccrrrrrrrr}
\caption{Raw circuit resource counts for all molecules, mappers, and tapering/frozen-core combinations. Columns show: number of spin orbitals ($N_{\text{so}}$), Hamiltonian terms ($N_{\text{sqh}}$), Pauli strings ($N_{\text{PS}}$), total Pauli operators ($N_{\text{PO}}$), and individual Pauli operator counts ($N_X$, $N_Y$, $N_Z$, $N_I$). Mapper abbreviations: JW = Jordan--Wigner, BK = Bravyi--Kitaev, P = Parity. Combo: (-) = no tapering/no frozen core (baseline), (T) = tapering only, (F) = frozen core only, (TF) = both.}\label{tab:raw_data_A}\\
\toprule
\textbf{Molecule} & \textbf{Mapper} & \textbf{Tap.} & \textbf{FC} & \textbf{Combo} & $N_{\text{so}}$ & $N_{\text{sqh}}$ & $N_{\text{PS}}$ & $N_{\text{PO}}$ & $N_X$ & $N_Y$ & $N_Z$ & $N_I$ \\
\midrule
\endfirsthead
\caption[]{(\textit{continued})}\\
\toprule
\textbf{Molecule} & \textbf{Mapper} & \textbf{Tap.} & \textbf{FC} & \textbf{Combo} & $N_{\text{so}}$ & $N_{\text{sqh}}$ & $N_{\text{PS}}$ & $N_{\text{PO}}$ & $N_X$ & $N_Y$ & $N_Z$ & $N_I$ \\
\midrule
\endhead
\midrule \multicolumn{13}{r}{\textit{Continued on next page}} \\
\endfoot
\bottomrule
\endlastfoot
\textit{H$_2$} & JW & \textbf{False} & \textbf{False} & \textbf{-} & \textbf{4} & \textbf{36} & \textbf{15} & \textbf{60} & \textbf{8} & \textbf{8} & \textbf{16} & \textbf{28} \\
 &  & True & False & T & 4 & 36 & 3 & 3 & 1 & 0 & 1 & 1 \\
 &  & False & True & F & 4 & 36 & 15 & 60 & 8 & 8 & 16 & 28 \\
 &  & True & True & TF & 4 & 36 & 3 & 3 & 1 & 0 & 1 & 1 \\
\cmidrule(lr){2-13}
 & BK & \textbf{False} & \textbf{False} & \textbf{-} & \textbf{4} & \textbf{36} & \textbf{15} & \textbf{60} & \textbf{8} & \textbf{0} & \textbf{26} & \textbf{26} \\
 &  & True & False & T & 4 & 36 & 3 & 3 & 1 & 0 & 1 & 1 \\
 &  & False & True & F & 4 & 36 & 15 & 60 & 8 & 0 & 26 & 26 \\
 &  & True & True & TF & 4 & 36 & 3 & 3 & 1 & 0 & 1 & 1 \\
\end{longtable}
\end{landscape}

\begin{landscape}
\begin{longtable}{llcccrrrr}
\caption{Raw circuit resource counts (continued) --- circuit width, depth, total gates, and number of variational parameters for all molecules, mappers, and combinations. (-) rows (bold) serve as the baseline for reduction calculations in Table~\ref{tab:reductions_data}. Mapper abbreviations and combo codes as in Table~\ref{tab:raw_data_A}.}\label{tab:raw_data_B}\\
\toprule
\textbf{Molecule} & \textbf{Mapper} & \textbf{Tap.} & \textbf{FC} & \textbf{Combo} & $W$ & $D$ & $G$ & $P$ \\
\midrule
\endfirsthead
\caption[]{(\textit{continued})}\\
\toprule
\textbf{Molecule} & \textbf{Mapper} & \textbf{Tap.} & \textbf{FC} & \textbf{Combo} & $W$ & $D$ & $G$ & $P$ \\
\midrule
\endhead
\midrule \multicolumn{9}{r}{\textit{Continued on next page}} \\
\endfoot
\bottomrule
\endlastfoot
\textit{H$_2$} & JW & \textbf{False} & \textbf{False} & \textbf{-} & \textbf{4} & \textbf{60} & \textbf{85} & \textbf{3} \\
 &  & True & False & T & 1 & 1 & 1 & 1 \\
 &  & False & True & F & 4 & 60 & 85 & 3 \\
 &  & True & True & TF & 1 & 1 & 1 & 1 \\
\cmidrule(lr){2-9}
 & BK & \textbf{False} & \textbf{False} & \textbf{-} & \textbf{4} & \textbf{54} & \textbf{66} & \textbf{3} \\
 &  & True & False & T & 1 & 2 & 2 & 1 \\
 &  & False & True & F & 4 & 54 & 66 & 3 \\
 &  & True & True & TF & 1 & 2 & 2 & 1 \\
\end{longtable}
\end{landscape}

\begin{landscape}
\begin{longtable}{llcccrrrrrrrrrrrr}
\caption{Reduction factors relative to the (-) (no tapering, no frozen core) baseline for each molecule, mapper, and combination. Values represent the ratio $v_{\text{-}} / v_{\text{combo}}$, so a value of $2.00$ indicates a $2\times$ reduction (halving of resource). Values below $1.00$ (red) indicate a marginal increase relative to baseline. (-) rows are the baseline and show $1.00$ by definition. Column headers: $N_{\text{so}}$ spin orbitals, $N_{\text{sqh}}$ Hamiltonian terms, $N_{\text{PS}}$ Pauli strings, $N_{\text{PO}}$ total Pauli operators, $N_X/N_Y/N_Z/N_I$ individual Pauli counts, $W$ circuit width, $D$ circuit depth, $G$ total gates, $P$ parameters. Dashes (--) indicate division by zero (e.g.\ $N_Y=0$ for H$_2$/Parity baseline).}\label{tab:reductions_data}\\
\toprule
\textbf{Molecule} & \textbf{Mapper} & \textbf{Tap.} & \textbf{FC} & \textbf{Combo} & $N_{\text{so}}$ & $N_{\text{sqh}}$ & $N_{\text{PS}}$ & $N_{\text{PO}}$ & $N_X$ & $N_Y$ & $N_Z$ & $N_I$ & $W$ & $D$ & $G$ & $P$ \\
\midrule
\endfirsthead
\caption[]{(\textit{continued})}\\
\toprule
\textbf{Molecule} & \textbf{Mapper} & \textbf{Tap.} & \textbf{FC} & \textbf{Combo} & $N_{\text{so}}$ & $N_{\text{sqh}}$ & $N_{\text{PS}}$ & $N_{\text{PO}}$ & $N_X$ & $N_Y$ & $N_Z$ & $N_I$ & $W$ & $D$ & $G$ & $P$ \\
\midrule
\endhead
\midrule \multicolumn{17}{r}{\textit{Continued on next page}} \\
\endfoot
\bottomrule
\endlastfoot
\textit{H$_2$} & JW & \textbf{False} & \textbf{False} & \textbf{-} & \textbf{1.00} & \textbf{1.00} & \textbf{1.00} & \textbf{1.00} & \textbf{1.00} & \textbf{1.00} & \textbf{1.00} & \textbf{1.00} & \textbf{1.00} & \textbf{1.00} & \textbf{1.00} & \textbf{1.00} \\
 &  & True & False & T & 1.00 & 1.00 & 5.00 & 20.00 & 8.00 & -- & 16.00 & 28.00 & 4.00 & 60.00 & 85.00 & 3.00 \\
 &  & False & True & F & 1.00 & 1.00 & 1.00 & 1.00 & 1.00 & 1.00 & 1.00 & 1.00 & 1.00 & 1.00 & 1.00 & 1.00 \\
 &  & True & True & TF & 1.00 & 1.00 & 5.00 & 20.00 & 8.00 & -- & 16.00 & 28.00 & 4.00 & 60.00 & 85.00 & 3.00 \\
\cmidrule(lr){2-17}
 & BK & \textbf{False} & \textbf{False} & \textbf{-} & \textbf{1.00} & \textbf{1.00} & \textbf{1.00} & \textbf{1.00} & \textbf{1.00} & \textbf{1.00} & \textbf{1.00} & \textbf{1.00} & \textbf{1.00} & \textbf{1.00} & \textbf{1.00} & \textbf{1.00} \\
 &  & True & False & T & 1.00 & 1.00 & 5.00 & 20.00 & 8.00 & -- & 26.00 & 26.00 & 4.00 & 27.00 & 33.00 & 3.00 \\
 &  & False & True & F & 1.00 & 1.00 & 1.00 & 1.00 & 1.00 & -- & 1.00 & 1.00 & 1.00 & 1.00 & 1.00 & 1.00 \\
 &  & True & True & TF & 1.00 & 1.00 & 5.00 & 20.00 & 8.00 & -- & 26.00 & 26.00 & 4.00 & 27.00 & 33.00 & 3.00 \\
\end{longtable}
\end{landscape}

\end{document}